\documentclass[a4paper,11pt]{article}
\usepackage{graphicx} 
\usepackage[title]{appendix}

\usepackage[utf8]{inputenc}
\usepackage[affil-sl,auth-sc]{authblk}
\usepackage{amssymb,amsmath,amsbsy}
\usepackage{bm}
\usepackage{appendix}
\usepackage[colorlinks=true,  linkcolor=blue, urlcolor=cyan, citecolor=cyan]{hyperref}%
\usepackage[top=1.3in, bottom=1.3in,left=0.9in,right=0.9in]{geometry}
\usepackage{cite}
\usepackage{float}
\usepackage{caption}
\usepackage{wasysym}
\usepackage{slashed}
\usepackage{indentfirst}
\usepackage{multirow}
\usepackage{feynmp-auto}
\usepackage{braket}
\usepackage[nottoc]{tocbibind} 
\usepackage{color}
\usepackage{array}
\usepackage{epstopdf} 
\usepackage{parskip}
\usepackage{comment}
\usepackage{cancel}

\newcommand{\marrow}[5]{%
    \fmfcmd{style_def marrow#1
    expr p = drawarrow subpath (1/4, 3/4) of p shifted 6 #2 withpen pencircle scaled 0.4;
    label.#3(btex #4 etex, point 0.4 of p shifted 14 #2);
    enddef;}
    \fmf{marrow#1,tension=0}{#5}}

\let\originalleft\left \let\originalright\right
\renewcommand{\left}{\mathopen{}\mathclose\bgroup\originalleft}
\renewcommand{\right}{\aftergroup\egroup\originalright}

\numberwithin{equation}{section}

\title{\bf Chiral Anomaly Cancellation and Neutral Triple Gauge Boson Vertices in the SM EFT}
\author{Dimitrios Beis\footnote{\tt d.beis@uoi.gr}~ and Athanasios Dedes\footnote{\tt adedes@uoi.gr}}
\affil{Division of Theoretical Physics, 
University of Ioannina, GR-45110, Greece }

\date{\today}

\begin{document}

\maketitle

\begin{abstract}
We demonstrate the cancellation of chiral anomalies in the Standard Model Effective Field Theory (SM EFT),  achieved through by a consistent choice of loop momentum routing in triangle diagrams with dimension-6 operator insertions. By enforcing gauge invariance and Bose symmetry, we show that Goldstone boson contributions cancel anomalies arising from massive gauge boson vertices, preserving the consistency of the SM EFT. We compute neutral triple gauge boson vertices at one loop, revealing dominant contributions from dimension-6 operators at all energies below the EFT cutoff. A UV-complete anomaly-free model with a heavy vector-like electron validates our approach, illustrating how heavy fermion decoupling generates SM EFT operators while maintaining anomaly cancellation. Our results highlight the phenomenological relevance of these vertices for probing new physics at colliders, particularly through dimension-6 effects that scale as the inverse of the centre of mass energy squared, $1/s$, offering a viable pathway for experimental detection.  
\end{abstract}

\newpage
\maketitle
{
  \hypersetup{linkcolor=black}
  \tableofcontents
}

\section{Introduction}

Chiral anomalies represent a profound insight into the interplay between quantum mechanics and symmetries in particle physics. An anomaly occurs when a symmetry present at the classical level of a theory fails to be preserved after quantization. Chiral anomalies specifically pertain to the non-conservation of chiral currents in the presence of gauge fields, which has crucial implications for the consistency of quantum field theories (QFTs)~\cite{Bell:1969ts,Adler:1969gk,Fujikawa:1979ay}. 
In the Standard Model (SM) of particle physics, anomaly cancellation is essential to ensure gauge invariance, unitarity and renormalizability. The SM achieves this through a delicate balance of the electromagnetic charges, the colour and the weak isospin representations of quarks and leptons within one generation~\cite{Bouchiat:1972iq,Gross:1972pv,KorthalsAltes:1972aq}. 

When extending the SM to include the effects of new physics at higher energy scales, we often use an effective field theory approach known as the Standard Model Effective Field Theory (SM EFT) (for reviews see, Refs.~\cite{Brivio:2017vri,Isidori:2023pyp}). 
This article discusses the relevance of chiral anomalies within the SM EFT, 
as well as their intimate connections to the triple gauge boson vertices.
SM EFT extends the SM by adding higher-dimensional operators formed from SM fields that respect the gauge symmetries of the SM. At low energies, the coefficients in front of these operators are suppressed by powers of a high-energy scale $\Lambda$, reflecting the effects of new physics that resides above this scale. The SM EFT Lagrangian can be expressed as:
\begin{equation}
\mathcal{L}_{\text{SM EFT}} = \mathcal{L}_{\text{SM}} + \sum_{d>4} \frac{C_{d}}{\Lambda^{d-4}} 
\mathcal{O}_{d}, 
\end{equation}
where $\mathcal{L}_{\text{SM}}$ is the SM Lagrangian, $\mathcal{O}_{d}$ represents the $d$-mass 
dimensional operators and $C_{d}$ the corresponding Wilson coefficients.
When extending the SM to SM EFT, the potential re-emergence of anomalies becomes a critical concern. The added higher-dimensional operators could, in principle, introduce new anomalies or modify existing ones. Therefore, constructing a consistent SM EFT demands careful attention to anomaly cancellation at each operator level.
If a Beyond the Standard Model (BSM) theory exists and is gauge anomaly-free, then the SM EFT should also be gauge anomaly-free in the sense that it preserves all Ward identities necessary for gauge invariance. In other words, it is imperative to match a consistent theory in the ultraviolet  with another consistent one in the infrared following the decoupling of heavy particles at the scale $\Lambda$~\cite{Appelquist:1974tg}.

A first idea that comes into one's mind is that cancellation of chiral anomalies happens al\'a SM and therefore
leads to sum rules among Wilson coefficients that should be obeyed in any anomaly-free fundamental (UV) theory 
above scale $\Lambda$~\cite{Cata:2020crs}. 
However, this procedure does not work: there are certain UV models invalidating 
these anomaly-cancellation sum rules~\cite{Bonnefoy:2020tyv}. 
The current state of the art is that chiral anomalies can be removed with an appropriate choice of
local and finite counterterms~\cite{Feruglio:2020kfq,Passarino:2021uxa,Cohen:2023gap}. 

Regarding chiral anomalies in the SM EFT,  our study serves as a complement to these works
for it follows a somewhat different direction. 
By working exclusively in the broken electroweak phase of the SM EFT~\cite{dedes:2017zog},
we calculate triangle one-loop diagrams strictly in four-dimensions with $\{\gamma^\mu,\gamma^5 \}=0$,
and demand Ward identities to be satisfied by a certain routing of the momenta in the triangle loop. 
This is certainly a scheme choice which is consistent with gauge invariance and Bose symmetry, 
as we readily confirm.
We prove that the procedure can always be performed with higher dimensional operator insertions
in SM EFT, the key points of its success being \textit{a)} the SM is an anomaly free gauge theory and \textit{b)} the Goldstone boson triangle diagrams cancelling massive vector-boson triangle diagrams in SM EFT.

We demonstrate the above within a bottom-up and a top-down approach and compare with the 
full theory result for a decoupling of a heavy vector-like fermion.
Our outcome in both approaches follows the picture found by D'Hoker and Farhi~\cite{DHoker:1984izu,DHoker:1984mif}:
the decoupling of a fermion whose mass is generated by a Yukawa coupling  results in a gauge invariant action functional with a gauge and Higgs field term. In our UV-considered
theory in section~\ref{sec:fulltheory}, the  vector-like electron which has a large gauge invariant
mass $M$, receives an additional small part from a Yukawa coupling $y_E$, proportional to 
$|y_E|^2 v^2/M$ as a result of the
mixing of heavy and light (SM) electrons after the Higgs field acquiring a vacuum expectation value $v$.
The induced terms are gauge invariant and are encoded in the SM EFT in agreement with the full theory.

Following the idea of Ref.~\cite{Dedes:2012me},\footnote{The same procedure has been applied
to the decay $Z\to Z'\gamma$ in Refs.~\cite{Michaels:2020fzj,Kribs:2022gri} 
for extending the SM with a $U(1)'$ symmetry
or a St\"uckelberg gauge boson. See also~\cite{Anastasopoulos:2006cz,Allanach:2018vjg,Anastasopoulos:2024bxx}.} 
we use the routing momenta parameters
consistent with the Ward Identities and Bose symmetry to calculate neutral Triple Gauge 
boson Vertices (nTGVs) in SM EFT with dimension-6 operator insertions
for all combinations $V^*VV$, with $V=\gamma,Z$. In doing so, we compare the results for 
the form factor $h_3^\gamma(s)$~\cite{Hagiwara:1986vm,Gounaris:1999kf} with a full (UV) theory  example
mentioned above, merely as a demonstration of the validity of our procedure. We find, and to our knowledge for the 
first time, that the  effects of dimension-6 operators~\cite{Grzadkowski:2010es}
\begin{eqnarray}
    \mathcal{O}^{(1)}_{\varphi \ell} &=& (\varphi^\dagger i \overset{\leftrightarrow}{D}_\mu \varphi) (\bar{\ell}_L \gamma^\mu \ell_L)\;, \quad 
    \mathcal{O}^{(3)}_{\varphi \ell} = (\varphi^\dagger i 
    \overset{\leftrightarrow}{D}{^{I}_\mu} \varphi) 
    (\bar{\ell}_L \tau^I \gamma^\mu \ell_L)\;, \quad \mathcal{O}_{\varphi e} = 
    (\varphi^\dagger i \overset{\leftrightarrow}{D}_\mu \varphi) (\bar{e}_R \gamma^\mu e_R)\;,
    \nonumber \\
    \mathcal{O}^{(1)}_{\varphi q} &=& (\varphi^\dagger i \overset{\leftrightarrow}{D}_\mu \varphi) (\bar{q}_L \gamma^\mu q_L)\;, \quad
    \mathcal{O}^{(3)}_{\varphi q} = (\varphi^\dagger i \overset{\leftrightarrow}{D}{^I_\mu} \varphi) (\bar{q}_L \tau^I \gamma^\mu q_L)\;, \quad
    \mathcal{O}_{\varphi d} = 
    (\varphi^\dagger i \overset{\leftrightarrow}{D}_\mu \varphi) (\bar{d}_R \gamma^\mu d_R)\;,
    \nonumber \\
    \mathcal{O}_{\varphi u} &=& 
    (\varphi^\dagger i \overset{\leftrightarrow}{D}_\mu \varphi) (\bar{u}_R \gamma^\mu u_R)\;,
    \label{eq:dim6ops}
\end{eqnarray}
are far more important than dimension-8 operator insertions~\cite{Murphy:2020rsh}, 
and for all energies below the New Physics (NP) resonances
scale like $1/s$ where $\sqrt{s}$ is the centre of mass energy.  
There is a promising region where
their effects could in principle be measured at future $e^+\,e^-$-colliders 
(subject to electroweak constraints and collider sensitivity) 
at energies above but nearby the electroweak scale. 
The LHC and its upgraded Hi-Luminosity (HL-LHC) setup also hold similar 
conclusions at high (TeV) energies,  despite the small size of the form factors.

Our study, as introduced above, is an interesting probe for collider phenomenology.
Since LEP~\cite{OPAL:2000ijr,L3:2004hlr,DELPHI:2007gzg}, searches for neutral triple gauge boson vertices have attracted sufficient attention, both experimentally at Tevatron~\cite{D0:2009olv,CDF:2011rqn} and LHC~\cite{ATLAS:2018nci,CMS:2016cbq,ATLAS:2019gey}
as well as theoretically, a partial list of references include~\cite{Barroso:1984re,Baur:1992cd,Dedes:2012me,Gounaris:2000tb,Alcaraz:2001nv,Cata:2013sva,Degrande:2013kka,Senol:2022snc,Ellis:2023ucy,Novales-Sanchez:2023ztg,Subba:2023jia,Jahedi:2023myu, Cepedello:2024ogz,Ellis:2024omd,Medina:2025ypm}.
Furthermore, applications of triple gauge boson vertices to dark matter~\cite{Dudas:2012pb,Dror:2017nsg,DAgnolo:2020mpt,Medina:2021ram}, 
neutron stars~\cite{Berryman:2022zic}, and cosmology~\cite{Clery:2023mjo} 
complement an interesting topic in the current literature.

The outline of our study goes as follows: in section~\ref{sec:WI} we define the relevant 
Ward Identities (WIs) for addressing chiral anomalies 
through the gauge-fixing parameter independence. In section~\ref{sec:AnomSMEFT}
we show how these WIs can be made satisfied in SM EFT 
with dimension-6 operator insertions. In section~\ref{sec:nTGC}
we discuss analytically the impact of EFT operators in nTGVs and in section~\ref{sec:topdown} we work 
through a particular UV-model to verify our SM EFT results and go beyond to discuss higher dimensional
operator effects. The reader who is solely interested in nTGVs can start directly 
from section~\ref{sec:nTGC} without
losing any content and can continue to section~\ref{sec:pheno} where phenomenological 
implications are briefly stated. We conclude in section~\ref{sec:conclusions}.
All the relevant vertices discussed in this article are included 
in the supplement material in Appendix~\ref{app:A}.

\section{A Ward Identity from gauge-fixing-parameter independence}
\label{sec:WI}

\begin{figure}
    \centering
\begin{fmffile}{Zboson}
\parbox{40mm}{
 \begin{fmfgraph*}(120,80)
\fmfleftn{i}{2}
\fmfrightn{o}{2}
\fmf{fermion}{i1,v1,i2}
\fmflabel{$e^-$}{i1}
\fmflabel{$e^+$}{i2}
\fmf{boson,label=${Z}\; \mathbf{\rightarrow}{q}$}{v1,v2}
\fmf{fermion,label=$f$,label.side=left,tension=0.4}{v3,v2}
\fmf{boson}{v3,o1}
\fmf{fermion,tension=0.4}{v4,v3}
\fmf{boson}{v4,o2}
\fmf{fermion,tension=0.4}{v2,v4}
\fmflabel{$V_j$}{o2}
\fmflabel{$V_k$}{o1}
\fmfv{d.shape=circle,d.size=0,l=$\sigma$,l.a=70}{v1}
\fmfv{d.shape=circle,d.size=0,l=$\mu$,l.a=110}{v2}
\fmfv{d.shape=circle,d.size=0,l=$\rho$,l.a=-110}{v3}
\fmfv{d.shape=circle,d.size=0,l=$\nu$,l.a=110}{v4}
\marrow{ea}{left}{top}{$p_2$}{i1,v1}
\marrow{eb}{left}{top}{$p_1$}{i2,v1}
\marrow{ec}{right}{bot}{$k_1$}{v4,o2}
\marrow{ed}{right}{top}{$k_2$}{v3,o1}
\end{fmfgraph*}}
  \quad + \quad
\parbox{40mm}{
 \begin{fmfgraph*}(120,80)
\fmfleftn{i}{2}
\fmfrightn{o}{2}
\fmf{fermion}{i1,v1,i2}
\fmf{dashes,label=$G^0$}{v1,v2}
\fmf{fermion,label=$f$,tension=0.4}{v3,v2}
\fmf{photon}{v3,o1}
\fmf{fermion,tension=0.4}{v4,v3}
\fmf{photon}{v4,o2}
\fmf{fermion,tension=0.4}{v2,v4}
\end{fmfgraph*}}
\quad + \quad {\large crossed diagrams}
\end{fmffile}
\vspace{0.5cm}
    \caption{\sl Feynman diagrams for $s$-channel contributions to $e^+ e^- \to V_j V_k$ 
    amplitude, where $V=(Z,W,\gamma)$ and mediators the $Z$-gauge boson and  
    Goldstone boson ($G^0$) triangle loop circulated by a fermion of flavour $f$.}
    \label{fig:1}
\end{figure}
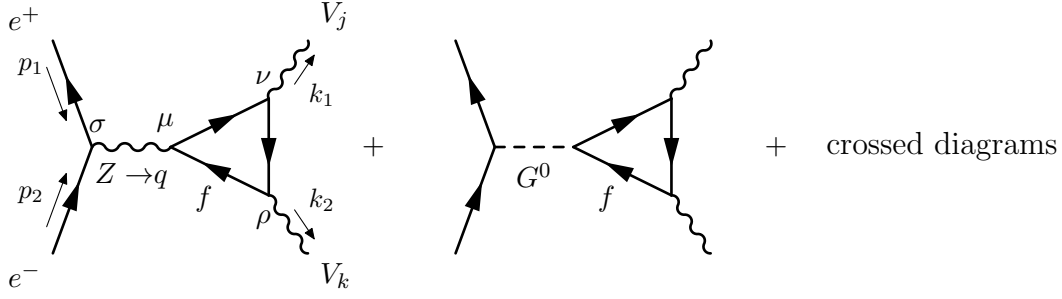

Let's first determine the relevant Ward-Identity (WI), and then show why it is important to be anomaly free.
As an example, we consider the $s$-channel of a positron-electron scattering process with momenta $p_1$ and $p_2$, respectively, to final state on-shell vector bosons, 
$V_j^\nu$ and $V_k^\rho$, with momenta $k_1$ and $k_2$, respectively, 
$e^{+}(p_1) + e^{-}(p_2) \to V_i^*(q)\to  V^\nu_j(k_1) + V^\rho_k(k_2)$  with $V$ being either the 
massive neutral $(Z)$, or charged  ($W^\pm$) gauge bosons
or the photon, $\gamma$. Momentum conservation implies $q_\mu=(k_1+k_2)_\mu=(p_1+p_2)_\mu$ 
and the Greek letters $\mu, \nu, \rho,...$ indicate
spacetime indices. Working in linear $R_\xi$-gauges,\footnote{We are working in Warsaw basis~\cite{Grzadkowski:2010es}, adopting the 
notation and the Feynman Rules in mass basis from Ref.~\cite{dedes:2017zog}. The Feynman Rule for the 
gauge boson ($V$) vertex to fermions ($f$) is expressed
as $[-i \gamma_\mu (a_{V}^{(f)} + b_V^{(f)} \gamma^5)]$, throughout. 
Their explicit form in SM EFT is provided in Appendix~\ref{app:A}.} 
the relevant Feynman diagrams are those with $\gamma$ or $Z$ mediators, the latter associated
with the corresponding Goldstone boson, $G^0$, depicted in Fig.~\ref{fig:1}.
The scattering amplitude, in standard textbook notation (e.g.~\cite{Peskin:1995ev}), reads,
\begin{align}
    &-\bar{v}(p_1) [ \gamma^\sigma (a^{(e)}_{Z} + b^{(e)}_{Z} \gamma^5)] u(p_2)\, \frac{1}{q^2- M_{Z}^2} 
    \left ( g_{\sigma \mu} - \frac{q_\sigma q_\mu}{M_{Z}^2} \right )\,\sum_f \Delta^{\mu\nu\rho\, (f)}_{Z V_j V_k} (k_1,k_2)
    \; \epsilon^*_\nu(k_1) \epsilon^*_\rho(k_2)
    \nonumber \\[2mm]
    &+ \frac{2\, b_{Z}^{(e)}\: m_e}{q^2-\xi_{Z} M_{Z}^2} \frac{1}{M_{Z}^2} \: \bar{v}(p_1) \gamma^5 u(p_2)\, 
    \sum_f \left [ q_\mu\, \Delta_{Z V_j V_k}^{\mu\nu\rho \, (f)}(k_1,k_2) 
     \ +\ i\: M_{Z} \, \Delta_{G^0V_jV_k}^{\nu \rho\, (f)}(k_1,k_2) \right ] \; \epsilon^*_\nu(k_1) \epsilon^*_\rho(k_2)
     \;.
    \label{eq:amp}
\end{align}
The one-particle irreducible (1PI) vertices, $\Delta^{\mu\nu\rho\, (f)}_{Z V_j V_k} (k_1,k_2)$ and $\Delta^{\nu\rho\, (f)}_{G^0 V_j V_k} (k_1,k_2)$, denote  the fermion-$f$ triangle vertices with the 
$Z-$ and Goldstone $G^0-$bosons in the $i$-th vertex, respectively.
Only the term in the second line of \eqref{eq:amp} depends on the gauge-fixing parameter, $\xi_Z$, through the 
pole in the Goldstone boson propagator. One can convince themselves that in the whole amplitude there is no other  contribution to such a term,\footnote{It is straightforward, albeit tedious, to demonstrate this result, 
but there is 
a quick way through using the three gauge fixing terms one for each vector boson propagator, 
$\xi_\gamma,\xi_Z$ and $\xi_{W^\pm}$. 
Since there is no other triangle or box diagram with \textit{all} fermions involved
mediated by, $Z$-boson (or the photon),
we arrive at the only two options for diagrams in Fig.~\ref{fig:1}.} 
and as a consequence, the expression inside the parenthesis must vanish. 
Keeping in mind current conservation for $V_i$, the Ward-Identity among on-shell gauge-bosons $V_j$ and $V_k$
states,\footnote{The outside product of polarization vectors is kept implicit in all the WIs below unless stated otherwise.} can be generalised as follows:
\begin{equation}
 q_\mu\, \Delta_{V_i V_j V_k}^{\mu\nu\rho} \ +\ i\: M_{V_i} \, \Delta_{G^0V_jV_k}^{\nu \rho}
 \ = \ 0\;,  
 \qquad V_i=Z,\gamma \;, \quad V_{j,k} = Z,\gamma, W^\pm \;,
 \label{eq:WI1}
\end{equation}
 where $\Delta_{V_i V_j V_k}^{\mu\nu\rho} = \sum_f \Delta_{V_i V_j V_k}^{\mu\nu\rho\, (f)}$  and 
 similarly, $\Delta_{G^0V_jV_k}^{\nu \rho}=\sum_f \Delta_{G^0V_jV_k}^{\nu \rho\, (f)}$, the
3-point 1PI triangle graphs shown in Fig.~\ref{fig:1} summed over all relevant fermions-$f$.
Note that the WI~\eqref{eq:WI1} is an important piece for  theory consistency, for if otherwise the 
scattering amplitude in \eqref{eq:amp} would depend on the gauge fixing parameter $\xi_Z$ (or $\xi_\gamma$
for a photon mediator instead). 
It is straightforward to observe  that equation \eqref{eq:WI1} is equivalent to the validity of 
partial conservation of axial-vector current at the classical level.

Because in the SM EFT there are two neutral weak bosons, $Z$ and $\gamma$, at least two out of three
gauge bosons should be identical, and Bose-symmetry should apply. For example, in Fig.~\ref{fig:1}, if
$V_j=Z$ and $V_k=\gamma$, the $S$-matrix element should be identical to the 
one with $q\leftrightarrow -k_1$. Considering this way, 
 there are analogous WIs to Eq.~\eqref{eq:WI1}, for the fields $V_j^\nu$ and $V_k^\rho$, namely,
\begin{align}
    -k_{1\nu} \, \Delta_{V_i V_j V_k}^{\mu\nu\rho} \ + \ i\: M_{V_j} \, \Delta_{V_iG^0V_k}^{\mu \rho} \ &= \ 0\;,
 \qquad V_j=Z,\gamma \;, \quad V_{i,k}=Z,\gamma, W^\pm \;,
    \label{eq:WI2}\\
    -k_{2\rho} \, \Delta_{V_i V_j V_k}^{\mu\nu\rho} \ + \ i\: M_{V_k} \, \Delta_{V_iV_jG^0}^{\mu \nu} \ &= \ 0\;,
 \qquad V_k=Z,\gamma \;, \quad V_{i,j}=Z,\gamma, W^\pm\;.
    \label{eq:WI3}
\end{align}
One more, but little laborious, way of deriving Eqs.~\eqref{eq:WI2} and \eqref{eq:WI3} is the following: 
stick a two fermion-current in every outgoing vector-boson of Fig.~\ref{fig:1} and inspect gauge invariance
in the sense of $\xi$-independence. Not only these Ward-Identities [together with Eq.~\eqref{eq:WI1}] are needed
for gauge invariance to hold, but also reduced Ward-Identities of the form 
\begin{equation}
    q_\mu \Delta_{V_i\dots}^{\mu\dots} + i M_{V_i} \Delta^{\, \dots}_{G_i\dots}=0\;,
    \label{eq:WIgeneral}
    \end{equation}
are in place, where the dots represent a combination of Goldstone and (or) vector boson fields. 
Therefore, Eqs.~\eqref{eq:WI1}\eqref{eq:WI2} and \eqref{eq:WI3} should always be valid at the same time.
Their violation renders the SM, and therefore the SM EFT, meaningless.

One can derive similar identities when working with charged currents too, e.g. similar to Fig.~\ref{fig:1} 
but with the process $\bar{\nu}_e(p_1) + e^-(p_2)\to V^*_i(q)  \to V_j(k_1) + V_k(k_2)$, 
where two $V$s  are $W$-gauge bosons. In fact, it is possible to group all the WIs in a single equation
for every \textit{incoming} gauge boson $V_i$ (associated with the Goldstone boson $G_i$) 
with four-momentum transfer $q=k_1+k_2$ and mass $M_{V_i}$,
\begin{equation}
    q_\mu \, \Delta^{\mu\nu\rho}_{V_i V_j V_k}(k_1,k_2) + \beta(V_i) \, M_{V_i} \, \Delta^{\nu\rho}_{G_i V_j V_k}(k_1,k_2) = 0\;, \quad \{ \beta(Z)=i\;,\; \beta(W^+)=-1\;,\; \beta(W^-)=+1 \}\;,
    \label{eq:WImaster}
\end{equation}
graphically pictured in Fig.~\ref{fig:WI}, where the shaded triangles also include the crossed diagrams.
\begin{figure}[t]
    \centering
\begin{fmffile}{WardIdentity}
  \parbox{30mm} {\begin{fmfgraph*}(50,50)
    \fmfstraight
    \fmfleft{i1}
    \fmfright{o1,o2}
    \fmf{photon, label=$\mathbf{\rightarrow} q$}{i1,v}
    \fmf{photon}{i1,v}
    \fmfv{decor.shape=triangle,decor.filled=shaded,
    decor.size=30,decor.angle=-35}{v}
    \fmf{photon}{v,o1}
    \fmf{photon}{v,o2}
    \fmflabel{$q_\mu \cdot \Biggl ( V_i^\mu$}{i1}
    \fmflabel{$V_k^\rho$}{o1}
    \fmflabel{$V_j^\nu$}{o2}
    \marrow{b}{right}{bot}{$k_1$}{v,o2}
    \marrow{c}{right}{bot}{$k_2$}{v,o1}
  \end{fmfgraph*}}
   $\Biggr )$ \quad + $\beta(V_i) \; M_{V_i} \Biggl ($\qquad  
    \parbox{30mm}{\begin{fmfgraph*}(50,50)
    \fmfstraight
    \fmfleft{i1}
    \fmfright{o1,o2}
        \fmf{dashes, label=$\mathbf{\rightarrow} q$}{i1,v}
    \fmfv{decor.shape=triangle,decor.filled=shaded,
    decor.size=30,decor.angle=-35}{v}
    \fmf{photon}{v,o1}
    \fmf{photon}{v,o2}
    \fmflabel{$G_i$}{i1}
    \fmflabel{$V_j^\nu$}{o2}
    \fmflabel{$V_k^\rho$}{o1}
    \marrow{b}{right}{bot}{$k_1$}{v,o2}
    \marrow{c}{right}{bot}{$k_2$}{v,o1}
  \end{fmfgraph*}}
    $\Biggr )$ \quad =  \quad {\large 0}
\end{fmffile}
\vspace{0.5cm}
    \caption{\sl Graphical representation of the WI in Eq.~\eqref{eq:WImaster}.}
    \label{fig:WI}
\end{figure}

However, the WI \eqref{eq:WImaster} is, in general, false, \textit{i.e. it may be anomalous}.
The reason being the fact that the triangle diagrams $\Delta_{V_i V_j V_k}^{\mu\nu\rho \, (f)}$ in SM and in SM EFT and,
 $\Delta_{G^0V_jV_k}^{\nu \rho \, (f)}$ only in SM EFT, are linearly divergent. The calculation of 
 these diagrams leads to anomalous terms that depend on the routing 
 of the momentum circulating in (both) triangle loops,
which reflects upon the ambiguity, in the dimensional regularization scheme, 
of treating $\gamma^5$ consistently in higher than four dimensions. 
%

Strictly speaking, the identity \eqref{eq:WImaster} is valid for on-shell particles $V_j$ and $V_k$.
As such, it has been proven to all orders by Chanowitz and Gaillard~\cite{Chanowitz:1985hj}. 
Naturally, as the reader may already be aware, Eq.~\eqref{eq:WImaster} serves as the foundation for the renowned Goldstone Boson Equivalence Theorem (GBET)~\cite{Cornwall:1973tb,Cornwall:1974km,Vayonakis:1976vz,Lee:1977eg}.

\section{Chiral Anomalies in the SM EFT}
\label{sec:AnomSMEFT}

\subsection{SM EFT at tree level and WIs}
\label{sec:tree}

Because of gauge invariance and Bose-symmetry, 
in the SM EFT with up-to dimension-6 operators, there are no 
triple neutral gauge boson vertices at tree-level. There 
are, however, triple  vertices with charged vector bosons of the form $\gamma W^+ W^-$
and $Z W^+ W^-$ and higher particle multiplicity vertices arising from one CP-even and two CP-odd operators.
We have explicitly checked the validity of Eq.~\eqref{eq:WImaster} for those vertices at tree level 
by using the Feynman Rules from Ref.~\cite{dedes:2017zog} in $R_\xi$-gauges.

Nevertheless, for inspecting chiral anomalies in triangle diagrams in SM EFT, the
application of \eqref{eq:WIgeneral} is very useful. It is trivial to show that,
dotting a $V\bar{f}f$-vertex for every fermion flavour in SM EFT with the 
momentum of a massive vector boson $V$, we obtain a vertex proportional to $G\bar{f}f$-vertex 
with the vector boson being replaced by its associate Goldstone boson $G$.

\subsection{Cancellations of the triangle anomalies in the SM EFT}
\label{sec:triangles}

The naive WIs \eqref{eq:WImaster} contain two parts: the 
first triangle $\Delta^{\mu\nu\rho}_{V_iV_jV_k}$ attached to 
gauge bosons $V$, and the second triangle, $\Delta^{\nu\rho}_{G_iV_jV_k}$
where $V_i$ has been replaced by a Goldstone boson. Both triangles and their crossed diagrams
are linearly divergent. This means that each amplitude depends on the routing of
the momenta circulating the triangles.

For the neutral\footnote{For external $W$-bosons, the triangle fermion mass $m_f$ is not universal, but 
the diagrams share the same anomaly factor [\textit{c.f.} \eqref{eq:anterms}]
as for the case of neutral external gauge bosons. As far as the chiral
anomaly factor  is concerned, the result is independent of $m_f$. See Ref.~\cite{Dedes:2012me} for the direct calculation in the scheme adopted here.} external vector-bosons-triangle, and for each fermion $f$, we have,

\bigskip
\begin{center}
\begin{fmffile}{tangles}
  \parbox{40mm}{\begin{fmfgraph*}(80,100)
    \fmfstraight
    \fmfleft{i1}
    \fmfright{o1,o2}
    \fmf{photon,label=$\rightarrow q$}{i1,v1}
    \fmf{fermion,tension=0.3,label=$p+k_1+a$,l.side=left}{v1,v2}
    \fmf{fermion,tension=0.3,label=$p+a$,l.side=left}{v2,v3}
    \fmf{fermion,tension=0.3,label=$p-k_2+a$,l.side=left}{v3,v1}
    \fmf{photon}{v2,o2}
    \fmf{photon}{v3,o1}
    \fmflabel{\vspace{10pt}$\mu$}{v1}
    \fmflabel{\vspace{-10pt}$\nu$}{v2}
    \fmflabel{\hspace{-10pt}$\rho$}{v3}
    \fmflabel{$V_i$}{i1}
    \fmflabel{$V_k$}{o1}
        \fmflabel{$V_j$}{o2}
            \marrow{ebb}{right}{bot}{$k_1$}{v2,o2}
     \marrow{ecc}{right}{top}{$k_2$}{v3,o1}
\end{fmfgraph*}}
 + {\large crossed diagram [$a\leftrightarrow b$, $k_1\leftrightarrow k_2$, $(\nu,\rho)\leftrightarrow (\rho,\nu)$]}
\end{fmffile}
\end{center}
\bigskip

\begin{align}
    \Delta^{\mu\nu\rho\, (f)}_{V_iV_jV_k}(k_1,k_2;a,b)=&  (-1)\, \mathrm{Tr} \,
    \int \frac{d^4p}{(2\pi)^4} [-i \gamma^\mu (a_{V_i}^{(f)} + b_{V_i}^{(f)} \gamma^5)] 
    \frac{i}{(\cancel{p}-\cancel{k}_2)+\cancel{a}-m_f} [-i \gamma^\rho (a_{V_k}^{(f)} + b_{V_k}^{(f)} \gamma^5)] \nonumber \\[2mm]
    &\times \qquad \frac{i}{\cancel{p}+\cancel{a}-m_f} [-i \gamma^\nu (a_{V_j}^{(f)} + b_{V_j}^{(f)} \gamma^5)]
    \frac{i}{(\cancel{p}+\cancel{k}_1) +\cancel{a}-m_f}
    \nonumber \\[2mm]
    & +  (-1)\, \mathrm{Tr} \,
    \int \frac{d^4p}{(2\pi)^4} [-i \gamma^\mu (a_{V_i}^{(f)} + b_{V_i}^{(f)} \gamma^5)] 
    \frac{i}{(\cancel{p}-\cancel{k}_1)+\cancel{b}-m_f} [-i \gamma^\nu (a_{V_j}^{(f)} + b_{V_j}^{(f)} \gamma^5)] \nonumber \\[2mm]
    &\times \qquad \frac{i}{\cancel{p}+\cancel{b}-m_f} [-i \gamma^\rho (a_{V_k}^{(f)} + b_{V_k}^{(f)} \gamma^5)]
    \frac{i}{(\cancel{p}+\cancel{k}_2) +\cancel{b}-m_f}\;,
    \label{eq:trigono}
\end{align}    
where $(-1)$ is the fermionic loop factor and $m_f$ is a common mass for the fermion $f$ inside the triangle loop.
In each of the two diagrams, we shift the internal loop momenta with arbitrary four-vectors, $a^\mu$ and $b^\mu$.
This shift with arbitrary vectors is not necessary for the triangle diagram $\Delta^{\mu\nu\rho}$ 
itself, which is convergent after all; it is nevertheless necessary for its divergence, $q_\mu \Delta^{\mu\nu\rho}$, that we are going to examine below [\textit{c.f.} Eq~\eqref{eq:trigono2}].
All vector ($a^{(f)}_{V_i}$) or axial-vector ($b^{(f)}_{V_i}$) couplings are general SM EFT couplings to all orders
of EFT expansion.\footnote{Note that 
possible tensor (dipole) SM EFT couplings do not contribute to the chiral anomalies~\cite{Gerstein:1969cx}. They vanish when dotted with $q_\mu$. Moreover, dipole Wilson coefficients, arise at one-loop order in a perturbative
decoupled UV-theory, and are therefore formally a two-loop effect in the closed fermion loop diagrams.
Our analysis is strictly at one-loop order in the SM EFT.} 
Dotted \eqref{eq:trigono} with $q_\mu$ and exploiting standard identities\footnote{The Appendix B of Ref.~\cite{Dedes:2012me} contains all the  details.} we find
\begin{equation}
    q_\mu\: \Delta^{\mu\nu\rho\, (f)}_{V_iV_jV_k}(k_1,k_2;a,b) \ = \ 
    - 2\: m_f \,b^{(f)}_{V_i} \,\Gamma_{V_jV_k}^{\; \nu\rho\, (f)}(k_1,k_2)
    \ + \ \Pi_{V_i V_j V_k}^{\; \nu \rho\, (f)}(k_1,k_2;a,b) \;.
    \label{eq:trigono2}
\end{equation}
The integral $\Gamma_{V_jV_k}^{\; \nu\rho\, (f)}$ is finite and is given in Eqs.~(B4) and (B5) of Ref.~\cite{Dedes:2012me} and because it cancels  when the Goldstone diagram is included as shown below, 
is not a concern for the anomaly.
The important piece is the integral $\Pi_{V_i V_j V_k}^{\; \nu \rho\, (f)}(k_1,k_2;a,b)$. It is 
divided in two parts: a chiral, i.e., $\gamma^5$-dependent and a non-chiral, i.e., $\gamma^5$-independent.
For the non-chiral part, the choice $b^\mu=-a^\mu$~\cite{Dedes:2012me,Weinberg:1996kr,Dreiner:2023yus} of the arbitrary vectors results in, $\Pi_{V_i V_j V_k}^{\; \nu \rho\, (f)}(k_1,k_2;a,-a)|_{\mathrm{non-chiral}}=0$ which is expected because there are no non-chiral anomalies. 
Following this choice, we are left with one arbitrary vector $a^\mu$. We shall return
to the explicit formula of $\Pi_{V_i V_j V_k}^{\; \nu \rho\, (f)}(k_1,k_2;a,-a)|_{\mathrm{chiral}}$ 
after examining the Goldstone boson diagram below.\footnote{Admittedly, the use of the arbitrary shift-vector technique, while well-suited to chiral anomalies, may be  cumbersome when applied to
a general amplitude. See, however, Ref.~\cite{Dedes:2012hf}.}

In the SM EFT, by replacing the vector boson $V_i$ with its associate Goldstone boson, $G_i$ 
and utilizing a general form from Eqs.~\eqref{eq:G0nunu}-\eqref{eq:G0dd} from the Appendix~\ref{app:A}, we have 

\bigskip
\begin{center}
\begin{fmffile}{Gangles}
  \parbox{40mm}{\begin{fmfgraph*}(80,100)
    \fmfstraight
    \fmfleft{i1}
    \fmfright{o1,o2}
    \fmf{dashes,label=$\rightarrow q$}{i1,v1}
    \fmf{fermion,tension=0.3,label=$p+k_1+\tilde{a}$,l.side=left}{v1,v2}
    \fmf{fermion,tension=0.3,label=$p+\tilde{a}$,l.side=left}{v2,v3}
    \fmf{fermion,tension=0.3,label=$p-k_2+\tilde{a}$,l.side=left}{v3,v1}
    \fmf{photon}{v2,o2}
    \fmf{photon}{v3,o1}
    \fmflabel{\vspace{-10pt}$\nu$}{v2}
    \fmflabel{\hspace{-10pt}$\rho$}{v3}
    \fmflabel{$G_i$}{i1}
    \fmflabel{$V_k$}{o1}
        \fmflabel{$V_j$}{o2}
            \marrow{ebb}{right}{bot}{$k_1$}{v2,o2}
     \marrow{ecc}{right}{top}{$k_2$}{v3,o1}
\end{fmfgraph*}}
 + {\large crossed diagram [$\tilde{a}\leftrightarrow \tilde{b}$, $k_1\leftrightarrow k_2$, $(\nu,\rho)\leftrightarrow (\rho,\nu)$]}
\end{fmffile}
\end{center}
\bigskip
%
\begin{align}
    \Delta^{\nu\rho\, (f)}_{G_iV_jV_k}(k_1,k_2;\widetilde{a},\widetilde{b})\ = \ &  i \, \left [(-1)^f\, 
    \frac{m_f}{v Z_{G^0}} \right ]\, \Gamma_{V_jV_k}^{\; \nu\rho\, (f)}(k_1,k_2)
    \ + \ \frac{i}{M_Z} \, q_\mu \, \Delta^{\mu\nu\rho\, (f)}_{\widetilde{V}_iV_jV_k}(k_1,k_2;
    \widetilde{a},\widetilde{b})\;,
    \label{eq:trigonoG}
\end{align}    
where $(-1)^f=+1$ for  $f=u,\nu$ and $(-1)^f=-1$ for $f=d,e$. Note that the arbitrary momentum shift vectors
$\widetilde{a}^\mu$ and $\widetilde{b}^\mu$ are now, in general, different from the vector boson triangle.
Following the same reasoning as previously, we choose, $\widetilde{b}^\mu=-\widetilde{a}^\mu$.
The factor $Z_{G^0}$ is a wave-function normalization of the Goldstone boson and is 
a function of Wilson coefficients and the Higgs vacuum expectation value, $v$ (see Appendix~\ref{app:A} and Refs.~\cite{dedes:2017zog,Dedes:2023zws} for definitions). Our calculation below shows it cancels
together with all other field redefinitions.
Anomaly terms are originated
from the second term in the RHS of \eqref{eq:trigonoG} with the difference w.r.t \eqref{eq:trigono}
that the coupling of $\widetilde{V}_i=Z$ to the fermions is a pure EFT (dimension-6) coupling, i.e., it depends solely 
on Wilson coefficients as it is shown in Appendix~\ref{app:A}.

As we saw in Section~\ref{sec:WI}, the sufficient, and necessary chiral anomaly-free conditions to get 
gauge-fixing parameter independence is Eq.~\eqref{eq:WI1} and its variants, Eqs.~\eqref{eq:WI2} and \eqref{eq:WI3}.
We now have all the ingredients to check upon these WIs.
For the WI \eqref{eq:WI1}, we combine Eqs.~\eqref{eq:trigono2} and \eqref{eq:trigonoG},
and, by making use of the $V_i=Z$-boson with mass $M_Z=\frac{1}{2}g_Z v Z_{G^0}$, and denoting that, $b^{(f)}_Z=\widehat{b}^{(f)}_Z+\widetilde{b}_Z^{(f)}$ with $\widehat{b}_Z^{(f)}=-(-1)^f\frac{g_Z}{4}$ being the SM part, and,
$\widetilde{b}_Z^{(f)}$ the dimension-6 insertion 
of the axial $Zf\bar{f}$ vertex (see Appendix~\ref{app:A}), we find without loss
of generality,\footnote{Although 
we have taken $V_i=Z$, this equation is valid as well for $V_i=\gamma$ with $M_\gamma=0$ and 
$\Pi^{\nu\rho\, (f)}_{\widetilde{\gamma}_iV_jV_k} =0$ since, apart from dipole operators,
there are no other $d= 6$ EFT insertions
to the photon-fermion-fermion vertex  (they can be absorbed in the redefinition
of the electric charge~\cite{Dedes:2023zws}).}
\begin{equation}
    q_\mu\, \Delta_{V_i V_j V_k}^{\mu\nu\rho}(k_1,k_2;a) 
    \ +\ i\: M_{V_i} \, \Delta_{G^0V_jV_k}^{\nu \rho}(k_1,k_2;\widetilde{a}) \ = \ \sum_f\: \left [
    \Pi^{\nu\rho\, (f)}_{V_iV_jV_k}(k_1,k_2;a) - \Pi^{\nu\rho\, (f)}_{\widetilde{V}_iV_jV_k}(k_1,k_2;\widetilde{a}) \right ] \;,  
    \label{eq:WII1}
\end{equation}
where, the “tilde”  in $\Pi^{\nu\rho\, (f)}_{\widetilde{V}_iV_jV_k}$ 
means a pure dimension-6 insertion of the vertex 
$V_i$ to fermions, whereas all other vertices without the “tilde” are completely general, SM+EFT type. 
Note that, Eq.~\eqref{eq:WII1} is exact for every fermion mass $m_f$ and the 
rescaling factor $Z_{G^0}$ has disappeared.
Moreover, $\Pi^{\nu\rho\, (f)}_{V_iV_jV_k}$ contains only the chiral (i.e., dependent on $\gamma^5$) part,
since, as we pointed out previously, we can always choose the non-chiral part of $\Pi^{\nu\rho\, (f)}_{{V}_iV_jV_k}$ to vanish by taking the opposite routing of momentum shifts between the triangle diagram and its crossed cousin. This leaves just two unknown arbitrary vectors, $a^\mu$ and $\widetilde{a}^\mu$ in \eqref{eq:WII1}, related to $\Pi^{\nu\rho\, (f)}_{V_iV_jV_k}$ and $\Pi^{\nu\rho\, (f)}_{\widetilde{V}_iV_jV_k}$, respectively.
These two vectors can only depend upon a linear combination of the external momenta $k_1$ and $k_2$, namely,
\begin{equation}
    a^\mu = z \, k_1^\mu + w\, k_2^\mu\;, \qquad \widetilde{a}^\mu = \widetilde{z} \, k_1^\mu + \widetilde{w}\, k_2^\mu\;,
    \label{eq:routing}
\end{equation}
with the set $\{w,z,\widetilde{w},\widetilde{z}\}$ being real and arbitrary numbers.
Therefore, the routing of the momenta needs two parameters $w,z$ for the vector boson diagrams and two 
parameters, $\widetilde{w},\widetilde{z}$ for the Goldstone boson diagrams. We will show that 
these parameters can be 
chosen as such that chiral anomalies do cancel in SM EFT \textit{for each} SM fermion-$f$ contribution. 
There is an exception to this rule, which is famous~\cite{Bouchiat:1972iq,Gross:1972pv,KorthalsAltes:1972aq}:
The SM couplings introduce chiral anomalies which cancel only if we add all leptons
and quarks for each generation.

It is now time to discuss the explicit form of the chiral anomaly pieces, $\Pi_{V_iV_jV_k}^{\nu\rho\, (f)}$.
Following the steps in Appendix B of Ref.~\cite{Dedes:2012me} (see also Refs~\cite{Deser:1970spa, Weinberg:1996kr,Dreiner:2023yus}), using an anti-commuting-$\gamma^5$ and traces of 
$\gamma$-matrices strictly in four dimensions,
and after a bit of algebra, the RHS 
of Eqs.~\eqref{eq:WI1}\eqref{eq:WI2} and \eqref{eq:WI3} which is zero, now becomes 
\begin{align}
    \sum_f \left [ \Pi_{V_iV_jV_k}^{\nu\rho\, (f)}(k_1,k_2;a) - \Pi_{\widetilde{V}_iV_jV_k}^{\nu\rho\, (f)}(k_1,k_2;\widetilde{a}) \right ] &= \frac{1}{4\pi^2}\,
    L^{\nu\rho} \sum_f \: \left [ \mathcal{A}^{(f)}_{V_iV_jV_k}(w-z) - \mathcal{A}^{(f)}_{\widetilde{V}_iV_jV_k} (\widetilde{w} - \widetilde{z}) \right ]\;, \label{eq:R1} \\
    \sum_f \left [\Pi_{V_iV_jV_k}^{\mu\rho\, (f)}(k_1,k_2;a) - \Pi_{{V}_i\widetilde{V}_jV_k}^{\mu\rho\, (f)}(k_1,k_2;\widetilde{a}) \right ] &= \frac{1}{4\pi^2}\,
    L^{\mu\rho} \sum_f \: \left [ \mathcal{A}^{(f)}_{V_iV_jV_k}(w-1) - \mathcal{A}^{(f)}_{{V}_i\widetilde{V}_jV_k} (\widetilde{w} - 1) \right ]\;, \label{eq:R2} \\
    \sum_f \left [\Pi_{V_iV_jV_k}^{\mu\nu\, (f)}(k_1,k_2;a) - \Pi_{{V}_iV_j\widetilde{V}_k}^{\mu\nu\, (f)}(k_1,k_2;\widetilde{a}) \right ] &= \frac{1}{4\pi^2}\,
    L^{\mu\nu} \sum_f \: \left [ \mathcal{A}^{(f)}_{V_iV_jV_k}(z+1) - \mathcal{A}^{(f)}_{{V}_iV_j\widetilde{V}_k} (\widetilde{z} +1) \right ]\;, \label{eq:R3} 
\end{align}
where  $L^{\mu\nu}=\epsilon^{\mu\nu\kappa\lambda} k_{1\kappa} k_{2\lambda}$ is a Lorentz kinematic factor
for the anomaly, and, $\mathcal{A}^{(f)}_{V_iV_jV_k}$ is the anomaly factor 
\begin{equation}
    \mathcal{A}_{V_iV_jV_k}^{(f)} = a^{(f)}_{V_i} a^{(f)}_{V_j} b^{(f)}_{V_k} + a^{(f)}_{V_i} b^{(f)}_{V_j} a^{(f)}_{V_k} + b^{(f)}_{V_i} a^{(f)}_{V_j} a^{(f)}_{V_k} + b^{(f)}_{V_i} b^{(f)}_{V_j} b^{(f)}_{V_k} \;,
    \label{eq:anterms}
\end{equation}
for each fermion-$f$, within one generation, circulating the triangles.
The first three terms in \eqref{eq:anterms} represent the Vector-Vector-Axial (VVA) chiral anomaly, 
whereas the last term
the Axial-Axial-Axial (AAA) one. The coefficients $a^{(f)}$ and $b^{(f)}$ can be expanded as
\begin{equation}
    a^{(f)}_{V_i} = \widehat{a}^{(f)}_{V_i} + \widetilde{a}^{(f)}_{V_i}\;, \qquad b^{(f)}_{V_i} = \widehat{b}^{(f)}_{V_i} + \widetilde{b}^{(f)}_{V_i}\;, 
    \label{eq:a}
\end{equation}
where $\widehat{a}^{(f)}_{V_i}$ is the vector part of the
SM-like\footnote{As explicitly shown in Appendix~\ref{app:A}, 
the word SM-like  here and below means the following:
In the SM EFT, even at dimension-6 truncation, we need to rescale the 
gauge couplings  by certain normalization constants e.g. $Z_g$, $Z_{g'}$, etc, that 
depend on flavour blind Wilson coefficients associated with operators such as $Q_{\varphi B}$ and $Q_{\varphi W}$~\cite{dedes:2017zog,Dedes:2023zws}. 
These rescaling cancels out in the anomaly cancellation condition, {\it c.f.} Eq.~\eqref{eq:afree} like in the SM.},  dimension-4 vertex with fermion-$f$ and vector boson $V_i$,
and, $\widetilde{a}^{(f)}_{V_i}$ the 
pure dimension-6 corresponding part. Explicitly, all these couplings are produced in Appendix~\ref{app:A}.
Similarly, for the Axial-Vector couplings $b^{(f)}_{V_i}$. We can simplify further
Eqs.~\eqref{eq:R1}\eqref{eq:R2} and \eqref{eq:R3} by noting from Eqs.~\eqref{eq:a} and \eqref{eq:anterms},
that $\mathcal{A}^{(f)}_{\widetilde{V}_iV_jV_k}=
\mathcal{A}_{V_iV_jV_k}^{(f)} - \mathcal{A}_{\widehat{V}_iV_jV_k}^{(f)}$. Consequently, 
for the chiral anomalies to cancel in the SM EFT, 
the RHS of Eqs.~\eqref{eq:R1}\eqref{eq:R2} and \eqref{eq:R3} must vanish, that is
\begin{align}
   \sum_f \mathcal{A}^{(f)}_{V_iV_jV_k} (w-z-\widetilde{w}+\widetilde{z}) \ + \ \sum_f \mathcal{A}^{(f)}_{\widehat{V}_iV_jV_k} (\widetilde{w}-\widetilde{z}) \ &= \ 0\;, \label{eq:C1} \\
  \sum_f \mathcal{A}^{(f)}_{V_iV_jV_k} (w-\widetilde{w}) \ + \ \sum_f \mathcal{A}^{(f)}_{{V}_i\widehat{V}_jV_k} (\widetilde{w}-1) \ &= \ 0\;, \label{eq:C2} \\
   \sum_f \mathcal{A}^{(f)}_{V_iV_jV_k} (z-\widetilde{z}) \ + \ \sum_f \mathcal{A}^{(f)}_{{V}_iV_j\widehat{V}_k} (\widetilde{z}+1) \ &= \ 0\;. \label{eq:C3}
\end{align}
We now observe that the first term in the LHS of Eqs.~\eqref{eq:C1}\eqref{eq:C2} and \eqref{eq:C3} always vanishes
by the choice 
\begin{equation}
    w=\widetilde{w}\;, \quad z=\widetilde{z} \;.
    \label{eq:one}
\end{equation}
This selection indicates equal routing of momenta for the gauge and Goldstone boson diagrams in Fig.~\ref{fig:1}. 
In other words, for such a choice, the Goldstone boson triangle, pure $1/\Lambda^2$ contribution, cancels the corresponding triangle vector boson contribution.
With \eqref{eq:one}, the RHS of the Ward-Identities in
Eqs.~\eqref{eq:WI1}\eqref{eq:WI2} and \eqref{eq:WI3}, becomes 
\begin{align}
      \sum_f \mathcal{A}^{(f)}_{\widehat{V}_iV_jV_k} ({w}-{z}) \ &= \ 0\;, \label{eq:CC1} \\
 \sum_f \mathcal{A}^{(f)}_{{V}_i\widehat{V}_jV_k} ({w}-1) \ &= \ 0\;, \label{eq:CC2} \\
  \sum_f \mathcal{A}^{(f)}_{{V}_iV_j\widehat{V}_k} ({z}+1) \ &= \ 0\;. \label{eq:CC3}
\end{align}
Every WI contains one distinct SM-like vertex in a different vertex of the triangles, but
the other two vertices, the uncharted ones, are generic SM EFT  vertices.
Note, however, that if we naively take $w=z=0$ so that the
first WI \eqref{eq:CC1} is satisfied, as is the case implied in Ref.~\cite{Bonnefoy:2020tyv}, 
the other two ones never do so in general. 

First, let us assume that \textit{all} vertices, but one we contract with the external momentum in the triangle, 
are SM-like. For these, there is no solution set for $w$ and $z$ such that
anomalies cancel from all Ward Identities, i.e. the system of \eqref{eq:CC1}-\eqref{eq:CC3} has no solution,
unless it is, $\mathcal{A}^{(f)}_{\widehat{V}_i\widehat{V}_j\widehat{V}_k }=0$. This is what happens
in the SM, but only  when we add contributions from all leptons ($\ell$) and quarks ($q$) 
within a generation and for each term in \eqref{eq:anterms},
\begin{equation}
    \sum_{f=\ell,q} \mathcal{A}^{(f)}_{\widehat{V}_i\widehat{V}_j\widehat{V}_k } \ = \ 0 \;.
    \label{eq:afree}
\end{equation}
However, what about the cancellation of chiral anomalies when the vertex contracted is SM-like and the others are of SM EFT type or, more broadly, when there are multiple higher-dimensional operators entering the triangle vertices? 

It is enlightening to consider in detail one term from the RHS of
$\mathcal{A}^{(f)}_{V_iV_jV_k}$ in Eq.~\eqref{eq:anterms},
say $a^{(f)}_{V_i} b^{(f)}_{V_j} a^{(f)}_{V_k} \equiv a b a$ for clarity but strictly in this order, and explain
the logic behind anomaly cancellations at every order in EFT expansion. Then anomaly terms [apart from the pure SM one, $\widehat{a}\widehat{b}\widehat{a}$ which vanishes because of \eqref{eq:afree}]
violating the three WIs found in Eqs.~\eqref{eq:CC1}\eqref{eq:CC2} and \eqref{eq:CC3},
can be summarized in the following table:

\begin{table}[h]
    \centering
    \begin{tabular}{cccc}
        $\widehat{a}\widetilde{b} \widehat{a}$ & $\widehat{a}\widehat{b} \widetilde{a}$ & $\widehat{a}\widetilde{b} \widetilde{a}$ & $\propto (w-z)$\\
        $\widetilde{a}\widehat{b} \widehat{a}$ & $\widehat{a}\widehat{b} \widetilde{a}$ & $\widetilde{a}\widehat{b} \widetilde{a}$ & $\propto (w-1)$ \\
     $\widehat{a}\widetilde{b} \widehat{a}$    & $\widetilde{a}\widehat{b} \widehat{a}$ & $\widetilde{a}\widetilde{b} \widehat{a}$ & $\propto (z+1)$\\
    \end{tabular}
\end{table}
The key point here is to notice that each term appears in at most
two different rows of this matrix and, as such, the system for $w$ and $z$ \textit{always} 
has a solution. 
For example, the $\widehat{a}\widetilde{b} \widehat{a}$
term exists only in the first and third rows, so we can
choose $w=z=-1$ to cancel it. Similarly, the $\widehat{a}\widehat{b} \widetilde{a}$ term exists only
in the first and second rows, so we can choose $w=z=1$ to cancel it and, finally, 
the $\widetilde{a}\widehat{b} \widehat{a}$ term exists only in the second and third rows, and
we can choose $w=1, z=-1$ to cancel it. Therefore, one by one, all $1/\Lambda^2$ anomalous terms cancel.
The third column contains, potentially, $1/\Lambda^4$ terms. However, they are all unique 
and they cancel by choosing $w$ and $z$  such that the terms in parentheses indicating next to them, vanishes.
Therefore, $1/\Lambda^4$ anomaly terms cancel too. Furthermore, there are no $1/\Lambda^6$-terms in the anomaly
expansion with dimension-6 operator insertions. This completes the proof for chiral anomaly cancellation
with \textit{all} insertions of $d=6$ operators.

\begin{figure}[t]
    \centering
    \includegraphics[width=0.5\linewidth]{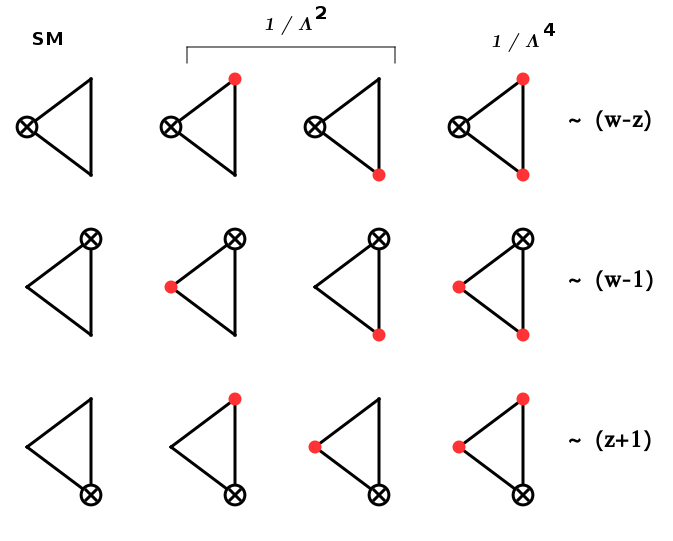}
    \caption{\sl The plan behind momentum routing choices for the pair of indices $(w,z)$ in four-vector 
    $a^\mu$ of Eq.~\eqref{eq:routing}
    so that chiral anomalies cancel in the SM EFT. 
    Every (red) blob indicates a pure EFT insertion, e.g. from $d=6$ operators only. The symbol “$\otimes$"
    means momentum action on the vertex indicated,  while non-dotted vertices are  SM-like vertices. 
    From top to bottom row, the triangles represent the three WIs in the SM EFT expansion, with anomaly terms in Eqs.~\eqref{eq:CC1}-\eqref{eq:CC3}.}
    \label{fig:2}
\end{figure}


One could attempt to elucidate upon the Eqs.~\eqref{eq:CC1}-\eqref{eq:CC3} somewhat graphically. 
In Fig.~\ref{fig:2} we have drawn the EFT expansion with $3\times 4=12$ triangles for the three
WIs. The symbol “$\otimes$” means dotted with external momentum which in turn means either cancellation
of Wilson coefficients in that vertex between massive vector and Goldstone bosons leaving behind a pure SM-vertex
or just a SM-like vertex. 
The red blob “\textcolor{red}{$\bullet$}” means an insertion of a dimension-6 operator.
A vertex without a symbol is a SM-like vertex. It is also helpful to refer to triangles 
in a matrix notation, e.g., (1,1) is the top-row, left-column triangle etc.
Again, it is more instructive to go through Fig.~\ref{fig:2} with 
an example. We adopt the triangle $\gamma Z Z$ (in order $ijk$) in SM EFT with dimension-6 operators.
A detailed expansion of the anomaly factors~\eqref{eq:anterms} corresponding to 
each of the three WIs~\eqref{eq:CC1}-\eqref{eq:CC3} and the use of Eq.~\eqref{eq:a}, are
\begin{align}
    \mathcal{A}_{\widehat{\gamma} Z Z}^{(f)} = \widehat{a}_\gamma^{(f)} b_Z^{(f)} a_Z^{(f)} + 
    \widehat{a}_\gamma^{(f)} a_Z^{(f)} b_Z^{(f)}
    \ &= \ \widehat{a}_\gamma^{(f)} \widehat{b}_Z^{(f)} \widehat{a}_Z^{(f)}
    + \widehat{a}_\gamma^{(f)} \widehat{a}_Z^{(f)} \widehat{b}_Z^{(f)} \nonumber \\
    &+ \widehat{a}_\gamma^{(f)} \widetilde{b}_Z^{(f)} \widehat{a}_Z^{(f)} 
     + \widehat{a}_\gamma^{(f)} \widehat{b}_Z^{(f)} \widetilde{a}_Z^{(f)}
     + \widehat{a}_\gamma^{(f)} \widetilde{a}_Z^{(f)} \widehat{b}_Z^{(f)} 
     + \widehat{a}_\gamma^{(f)} \widehat{a}_Z^{(f)} \widetilde{b}_Z^{(f)}\nonumber \\
    &+
    \widehat{a}_\gamma^{(f)} \widetilde{b}_Z^{(f)} \widetilde{a}_Z^{(f)} 
    +\widehat{a}_\gamma^{(f)} \widetilde{a}_Z^{(f)} \widetilde{b}_Z^{(f)}\;, 
    \label{eq:A1}\\[2mm]
    \mathcal{A}_{\gamma \widehat{Z} Z}^{(f)} = 
    a_\gamma^{(f)} \widehat{b}_Z^{(f)} a_Z^{(f)} + a_\gamma^{(f)} \widehat{a}_Z^{(f)} b_Z^{(f)}
    \ &= \ \widehat{a}_\gamma^{(f)} \widehat{b}_Z^{(f)} \widehat{a}_Z^{(f)}
    + \widehat{a}_\gamma^{(f)} \widehat{a}_Z^{(f)} \widehat{b}_Z^{(f)}\nonumber \\
    &+ \widehat{a}_\gamma^{(f)} \widehat{b}_Z^{(f)} \widetilde{a}_Z^{(f)}
     + \widehat{a}_\gamma^{(f)} \widehat{a}_Z^{(f)} \widetilde{b}_Z^{(f)}
    \;,\label{eq:A2} \\[2mm]
    \mathcal{A}_{\gamma Z \widehat{Z} }^{(f)} = 
    a_\gamma^{(f)} b_Z^{(f)} \widehat{a}_Z^{(f)} + a_\gamma^{(f)} a_Z^{(f)} \widehat{b}_Z^{(f)}
    \ &= \ \widehat{a}_\gamma^{(f)} \widehat{b}_Z^{(f)} \widehat{a}_Z^{(f)}
    + \widehat{a}_\gamma^{(f)} \widehat{a}_Z^{(f)} \widehat{b}_Z^{(f)}\nonumber \\
    &+ \widehat{a}_\gamma^{(f)} \widetilde{b}_Z^{(f)} \widehat{a}_Z^{(f)}
     + \widehat{a}_\gamma^{(f)} \widetilde{a}_Z^{(f)} \widehat{b}_Z^{(f)}
    \;. \label{eq:A3}
\end{align}
The photon vertex to fermions is purely vectorial and there is no
additional EFT insertion apart from universal redefinitions of the gauge couplings.
Therefore, triangles (2,2), (2,4), (3,3) and (3,4) do not exist for $\gamma ZZ$-triangle.
The first line of these equations corresponds to
the left column (1,1), (2,1), (3,1) of triangles in Fig.~\ref{fig:2}. 
They all cancel (for every $w$ and $z$) because
of \eqref{eq:afree}. The second line corresponds to $1/\Lambda^2$ contributions. The 
first and third terms in the second line of \eqref{eq:A1} [triangle (1,2) in Fig.~\ref{fig:2}]
exist only in the second line of \eqref{eq:A3} [triangle (3,2) in Fig.~\ref{fig:2}] 
and they cancel if we set the routing of momenta to $w=z=-1$ or, equivalently, the 
routing of the arbitrary vector $a^\mu = -k_1^\mu - k_2^\mu$. Similarly,
the second and fourth terms in the second line of \eqref{eq:A1} [triangle (1,3) in Fig.~\ref{fig:2}]
exist only in the second line of \eqref{eq:A2} [triangle (2,3) in Fig.~\ref{fig:2}] 
and they cancel if we set the routing of momenta to $w=z=1$ or, equivalently, the 
routing of the arbitrary vector $a^\mu = k_1^\mu + k_2^\mu$. Finally, the terms in the
third line of \eqref{eq:A1} are of order $1/\Lambda^4$ and
correspond to the triangle (1,4) in Fig.~\ref{fig:2}. They vanish automatically because we
set $w=z$ in the previous cases. To persuade oneself, one may perform additional examples involving all other external vector boson combinations.
 
The above examples, and other examples that the reader can come up with, have something in common.
The rather strange picture of the choice of the routing parameters, $w$ and $z$,
applies only when we look at SM vertices.
For instance, in the $\widehat{a}\widetilde{b}\widehat{a}$, we only need to satisfy the
WIs for the currents associated to SM-vertices insertions, $w=z$ and $z=-1$. 
For $\widetilde{a}\widetilde{b}\widehat{a}$ only one WI has to be satisfied,
the one associated with the $k$-leg, and therefore we choose $z=-1$. 
For three EFT insertions, say $\widetilde{a}\widetilde{b}\widetilde{a}$, 
there is no WI to satisfy, and therefore no anomalies are associated with them. It is the 
same with all $V_iV_jV_k$-triangles, as the reader can check by themselves. 
But what does this routing prescription imply at the end?
It implies \textit{that  divergence of currents arising from EFT insertions 
do not really matter at low energies regarding chiral anomalies.} 
As we shall see  below Eq.~\eqref{eq:LBSM} within a UV-model, these currents 
are associated with global symmetries that are not conserved anyway.
In other words, chiral anomalies arising from EFT insertions are rather illusory in nature. 

The freedom to choose arbitrary vectors $a^\mu$ and $b^\mu$  is thus
two-fold.
Firstly, the condition $b^\mu=-a^\mu$ eliminates the non-chiral anomaly,  and 
secondly, our choice of $(w,z)$ in $a^\mu = z \, k_1^\mu + w\, k_2^\mu$ as in picture Fig.~\ref{fig:2} moves  
the anomaly with dimension-6 SM EFT couplings to those currents with 
SM-like couplings. The latter anomaly cancels out when all the SM fermion fields add up,
\textit{i.e.} it is so because the SM is an anomaly free gauge theory.

Our analysis truncates the EFT expansion at dimension six, and we do not include a complete basis of dimension-eight operators. Nevertheless, certain contributions that scale as up-to $1/\Lambda^4$ can arise even within a dimension-six analysis. In this case, our choice for the pair $(w,z)$ for anomaly cancellation of $1/\Lambda^2$ insertions displayed in Fig.~\ref{fig:2} and highlighted by the
examples above, guarantees anomaly cancellation for the $1/\Lambda^4$ anomalous terms too.
Though we lack a proof, we are tempted to think that the same pattern will be repeated
for higher dimensional bases of operators.

In summary,  the disappearance of triangle chiral anomalies  in SM EFT  takes place term by term in Eq.~\eqref{eq:anterms}. The reason for cancellation is twofold:
first, because the SM is an anomaly free theory [e.g. Eq.~\eqref{eq:afree}] 
and second, because it is always possible 
to make a choice of the routing momenta for every SM fermion ($f$) circulating the loop triangle
when there is one or more insertions of higher dimensional  operators.

\subsection{Bose symmetry and chiral anomalies in the SM EFT}
\label{sec:Bose}
    
Is the routing of the momenta cancelling the anomalies consistent with Bose symmetry? The answer is yes.
This will be demonstrated here through several examples. 
Bose symmetry and anomaly cancellations for three identical gauge bosons imply 
the momentum routing parameters $z$ and $w$ which are displayed in Table~\ref{tab:1}.
%
\begin{table}
    \centering
    \begin{tabular}{|c|c|c|}
    \hline
       External legs & Bose symmetry & Anomaly cancellation\\ \hline
       $j\leftrightarrow k$  & $w+z=0$ & $w=1, z=-1$ \\ \hline
       $i\leftrightarrow j$  & $2w -z-1=0$ & $w=z=1$ \\ \hline
       $i\leftrightarrow k$  & $2z - w +1 =0$ & $w=z=-1$ \\ \hline
    \end{tabular}
    \caption{\sl Bose symmetry   
    and chiral anomaly cancellations routing momentum rules
    with a single operator dimension-6 insertion, among external, identical gauge-bosons.}
    \label{tab:1}
\end{table}
First, consider the example “$aba$” given below Eq.~\eqref{eq:afree} which corresponds 
to one of the various contributions in a $ZZZ$-triangle vertex. Then Bose symmetry applies among
identical bosons in SM or among identical bosons in SM EFT. For example, the term, $\widetilde{a}\widehat{b}\widehat{a}$ has a Bose symmetry among $j$ and $k$ external legs. Looking 
at Table~\ref{tab:1}, Bose symmetry is satisfied with $w+z=0$ and the cancellation of anomalies (with a single
dimension-6 insertion) by 
$w=1, z=-1$. Another example is the term $\widehat{a}\widetilde{b}\widehat{a}$ which 
exhibits a symmetry between $i$ and $k$ legs. For anomalies to cancel, we need, $w=z=-1$ which 
satisfies the requirement $2z-w+1=0$ for Bose symmetry. 
 Another example, not considered so 
far, is the $bbb$-term in $ZZZ$-triangle. Take for instance the $\widehat{b}\widehat{b}\widetilde{b}$
[Axial-Axial-Axial (AAA)] anomaly term. This is Bose symmetric under $i$ and $j$ legs, which from 
Table~\ref{tab:1} means $2w-z-1=0$ and the anomaly cancels when $w=z=1$.
Hence, in $ZZZ$ triangle, Bose symmetry and the cancellation of anomalies in SM EFT are inextricably linked.

Moving to a few more examples, we first consider 
the $\gamma ZZ$-vertex, which was worked out in Eqs.~\eqref{eq:A1}-\eqref{eq:A3},
and we consider for example the $\widehat{a}\widetilde{b}\widehat{a}$-term. This has a Bose-symmetry after
interchanging $j$ and $k$-legs in full SM EFT vertex, 
but has no Bose symmetry between the EFT insertion ($\widetilde{b}$)
and SM ($\widehat{a}$) vertices. Bose symmetry does not apply here. In contrary, for the vertex with
two EFT insertions, e.g.  
$\widehat{a}\widetilde{b}\widetilde{a}$, which is of order $(1/\Lambda^4)$,
we need $w+z=0$ for Bose-symmetry and $w-z=0$ for anomaly cancellation and are both satisfied when $w=z=0$.
Similarly, for the $Z\gamma Z$-triangle vertex, we have Bose symmetry among $i$ and $k$ legs
for identical $Z$-bosons which from Table~\ref{tab:1} implies $2 z -w +1=0$ and the anomaly 
cancellation in our prescription, e.g. $\widetilde{b}_Z \widehat{a} \widetilde{a}$ requires
$w=1$. Both constraints are fulfilled when $w=1$ (anomaly cancellation) and $z=0$ (Bose symmetry). 
All the above results for Bose symmetries
in triangle diagrams are in agreement with those found in Ref.~\cite{Dedes:2012me}.

 We should note in passing 
 that our scheme here is possibly not unique, since there may be other choices
 of the arbitrary vectors' combinations (including those associated with the Goldstone boson) 
 that preserve the WIs. 
We have tried several scheme choices for arbitrary vectors 
based solely on Bose symmetry [even relaxing  the constraint of Eq.~\eqref{eq:one}]
but did not succeed to find a simpler, meaningful picture when trying to 
preserve also the WIs.

 Nevertheless, 
the important conclusion is that, Bose-symmetry is \textit{compatible} with our prescription
of choosing the routing of the momenta for anomaly cancellation (see Table~\ref{tab:1}). 
Although Bose-symmetry alone, i.e. without gauge symmetry, 
is not sufficient to resolve the chiral anomaly cancellation in SM EFT,
determining the triple neutral gauge boson vertices requires Bose symmetry, 
as we will see in the next section.

\section{Neutral triple gauge boson triangle vertices}
\label{sec:nTGC}

Based on the previous analysis, we conclude that the $\xi$-parameter gauge dependence
in our working amplitude, $e^+e^-\rightarrow V_jV_k$ [see Eq.~\eqref{eq:amp} and Fig.~\ref{fig:1}], vanishes
for all combinations of SM EFT dimension-6 vertex insertions in the triangle. This is  a result of
vanishing chiral anomalies in SM EFT with a certain routing of momenta in different SM EFT triangle diagrams.
The master WI, Eq.~\eqref{eq:WImaster} can always be rendered to hold.

It is certainly of  phenomenological interest to check upon triple gauge boson vertices (TGVs) in the SM EFT
with dimension-6 insertions. Especially, neutral TGVs (nTGVs) are of special interest since they do not exist
at tree level in SM EFT in order $1/\Lambda^2$. 
We would like here to 
investigate whether,  finite threshold one-loop corrections appear from the triangle loop.
To that end, we write the most general  $C$-odd and $P$-odd,  hence CP-even,
vertex for three external gauge bosons, in the notation of the graph above
Eq.~\eqref{eq:trigono} [or Fig.~\ref{fig:1}], in the form~\cite{Dedes:2012me}\footnote{Apart from the appearance  
of parameters, $w$ and $z$, the expression \eqref{eq:Delta} utilized in an old work by Rosenberg~\cite{Rosenberg:1962pp}. Actually, there are two more form factors allowed in this expression, but they are reduced to \eqref{eq:Delta} after using certain identities; see footnote 11 of \cite{Dedes:2012me}.}
\begin{align}
    \Delta^{\mu\nu\rho}_{V_iV_jV_k}(k_1,k_2;w,z) &= A_1(k_1,k_2;w) \epsilon^{\mu\nu\rho\alpha} k_{2\alpha} 
    + A_2(k_1,k_2;z) \epsilon^{\mu\nu\rho\alpha} k_{1\alpha} 
    + A_3(k_1,k_2)\epsilon^{\mu\rho\alpha\beta}k_{2}^\nu k_{1\alpha}k_{2\beta} \nonumber \\[2mm]
    &+ A_4(k_1,k_2)\epsilon^{\mu\rho\alpha\beta}k_1^\nu k_{1\alpha}k_{2\beta}
    +   A_5(k_1,k_2)\epsilon^{\mu\nu\alpha\beta}k_2^\rho k_{1\alpha}k_{2\beta} \nonumber \\[2mm]
    &+   A_6(k_1,k_2)\epsilon^{\mu\nu\alpha\beta}k_1^\rho k_{1\alpha}k_{2\beta}\;.
    \label{eq:Delta}
\end{align}
The Lorentz invariant vertex $\Delta^{\mu\nu\rho}$ has mass dimension $+1$ 
and parametrizes the
result of the fermion triangle diagram.
Hence, $A_{1,2}$ are dimensionless and by naive power counting at most linearly divergent, and
therefore they depend on the regularization scheme or in our case, the routing momentum parameters $w,z$.
The form factors, $A_{3,\dots,6}$ on the other hand, have mass dimension $-2$ and are consequently finite; they 
can be calculated from the fermion triangle 
in any regularization scheme. The trick made in \cite{Dedes:2012me} is to apply 
\eqref{eq:Delta} in the already calculated Ward identities of 
Eqs.~\eqref{eq:WI2} and \eqref{eq:WI3} \textit{together with the anomaly terms}
of Eqs.~\eqref{eq:R2} and \eqref{eq:R3} in the RHS, and find $A_1(k_1,k_2;w)$ and $A_2(k_1,k_2;z)$ of \eqref{eq:Delta}. The result is then expressed in terms of the finite integrals $A_{3\dots 6}$ 
plus anomalous terms proportional to $w$ and $z$ which are defined from the anomaly cancellation conditions
and Bose symmetry described in the previous section.

As an example, the result for the triple neutral vertex $V_i^* V_j V_k$, where $V_i^*$ is an off-shell gauge boson
and $V_{j}$ and $V_k$ are on-shell vector bosons $V=\gamma,Z$, in momentum notation provided in Eq.~\eqref{eq:trigono},\footnote{Both sides of Eq.~\eqref{eq:mastervertex} are multiplied by the
product of polarization vectors $\epsilon_{\nu}^*(k_1) \epsilon_{\rho}^*(k_2)$, omitted here for clarity.} is
\begin{eqnarray}
    \Delta_{V_i^* V_j V_k}^{\mu\nu\rho}(k_1,k_2; w, z)  =  \sum_f
     & \Biggl \{\left [ k_1^2 (A_4^{(f)} - A_3^{(f)}) - \frac{m_f^2 b^{(f)}_{V_j}}{\pi^2} \, 
    I^{(f)}_{1\, V_i V_k} 
    + \frac{\mathcal{A}^{(f)}_{V_i{V}_jV_k}}{4\pi^2} (w-1) \right ] \, \epsilon^{\mu\nu\rho\alpha} k_{2\alpha} 
     \nonumber \\
    +  &\left [ k_2^2 (A_5^{(f)} + A_3^{(f)}) - \frac{m_f^2 b^{(f)}_{V_k}}{\pi^2} \, 
    I^{(f)}_{2\, V_i V_j}
    + \frac{\mathcal{A}^{(f)}_{V_iV_j{V}_k}}{4\pi^2} (z+1)\right ] \, \epsilon^{\mu\nu\rho\alpha} k_{1\alpha} 
     \nonumber \\
    + & A_3^{(f)} \, q^\mu \epsilon^{\nu\rho\alpha\beta} k_{1\alpha} k_{2\beta} \Biggr \}
     \;.
    \label{eq:mastervertex}
\end{eqnarray}
The integrals $A^{(f)}_{3,4,5}(k_1,k_2)$ and $I^{(f)}_{1,2}(k_1,k_2)$ are finite 
i.e., independent of $w$ and $z$, and their explicit expressions
together with several asymptotic limits and identities, can be found in Ref.~\cite{Dedes:2012me}.\footnote{The only notational difference is the anomaly factor denoted here as, $\mathcal{A}_{V_iV_jV_k}^{(f)}$ whereas in Ref.~\cite{Dedes:2012me} as $c$.}
The crucial terms in \eqref{eq:mastervertex} are the anomaly terms proportional to the routing
parameters $(w-1)$ and $(z+1)$: \textit{these must be taken precisely as 
described above in sections~\ref{sec:triangles} and~\ref{sec:Bose}
following the pattern indicated in Fig.~\ref{fig:2},
so that anomaly cancellation and Bose symmetry, and therefore gauge invariance, is preserved.}
In the following, we are going to apply Eq.~\eqref{eq:mastervertex} in deriving the neutral gauge 
boson form factors in the SM EFT with the dimension-6 operator insertions of \eqref{eq:dim6ops}.

\subsection{The $Z^*\gamma\gamma$ vertex}

We apply Bose symmetry between the two photons ($j$ and $k$ legs), and by setting $k_1^2=k_2^2=0$ and 
$b^{(f)}_{V_j}=b^{(f)}_{V_k}=0$ in Eq.~\eqref{eq:mastervertex} we find that 
\begin{equation}
    \Delta_{Z^*\gamma\gamma}^{\mu\nu\rho} (k_1,k_2;w,z) = \Delta_{Z^*\gamma\gamma}^{\mu\rho\nu} (k_2,k_1;w,z)
    \Longrightarrow w+z=0\;\; \mathrm{and} \;\; A_3^{(f)}(k_1,k_2)=A_3^{(f)}(k_2,k_1)\;.
    \label{eq:bose1}
\end{equation}
By inspecting the anomaly factor, 
\begin{equation}
    \sum_f \mathcal{A}_{Z\gamma\gamma}^{(f)} = \sum_f b_Z^{(f)} a_\gamma^{(f)} a_\gamma^{(f)} = 
    \sum_f 
    \widehat{b}_Z^{(f)} \widehat{a}_\gamma^{(f)} 
    \widehat{a}_\gamma^{(f)} + \sum_f \widetilde{b}_Z^{(f)} \widehat{a}_\gamma^{(f)} 
    \widehat{a}_\gamma^{(f)} \;,  \label{eq:azgg}
\end{equation}
the first term cancels because SM is an anomaly free theory and the second term because
of the choice $w=-z=1$ for anomaly cancellation in the SM EFT following the pattern of Fig.~\ref{fig:2}
and therefore, in accordance with Bose symmetry of Eq.~\eqref{eq:bose1}. 
The only term remaining from the RHS of Eq.~\eqref{eq:mastervertex} is the last one, proportional to $q^\mu$. 
This is also compatible with the Landau~\cite{Landau:1948kw} and Yang~\cite{Yang:1950rg} theorem, 
which states that the vertex $Z\gamma\gamma$
must vanish when all particles are on-shell due to invariance under rotation and inversion symmetries.
Looking at the first line of Eq.~\eqref{eq:amp}, the
$(s-M_Z^2)$-pole cancels, resulting in the amplitude
\begin{equation}
\sum_f \frac{2 m_e b_Z^{(e)}}{M_Z^2}\, A_{3\, Z^*\gamma\gamma}^{(f)}(s)\, 
[\bar{v}(p_1)\gamma^5 u(p_2)]  \, \epsilon^{\rho\nu\alpha\beta} k_{1\alpha} k_{2\beta}
\, \epsilon_{\nu}^*(k_1) \epsilon_{\rho}^*(k_2)\;,
\end{equation}
where $s=(k_1+k_2)^2$ is the centre of mass energy squared. 
Interestingly, in the limit of $s>>m_t^2$, the integral $A_3$ is approximated
with $\sum_f A_3^{(f)}(s>>m_t^2) \approx \sum_f \mathcal{A}_{Z\gamma\gamma}/(2\pi^2 s)$. 
Because of \eqref{eq:afree}, this term vanishes in the SM up to $1/s$, but not in the SM EFT. 
The conclusion is that, there are threshold one-loop corrections,
of the order $\frac{g^2}{4\pi^2}\frac{m_e^2}{M_Z^2}\frac{C^{(6)} v^2}{\Lambda^2}$,
contributing to the $s$-channel at high-energy  from the triangle diagram in the SM EFT with dimension-6 operators from the list of \eqref{eq:dim6ops} with Wilson coefficients collectively denoted here as $C^{(6)}$.
This  picture reveals the existence of heavy fermions, like the heavy electron below in section~\ref{sec:topdown}, and/or vector-bosons decoupled from UV-physics at low energies. However, 
their effects in $Z^*\gamma\gamma$ are utterly invisible at current colliders' architecture due to the external SM fermion mass suppression.
In the energy region below the top threshold,
$M_Z<s<m_t$, the SM anomaly cancellation $\sum_{f\ne t} \mathcal{A}^{(f)}_{Z\gamma\gamma} \ne 0$ is incomplete
and $A^{(f\ne t)}_3\approx -\frac{\mathcal{A}^{(f\ne t)}_{Z\gamma\gamma}}{24\pi^2 m_t^2} $. However, 
the contribution from
dimension-6 operators is further 
suppressed relative to the SM by a factor $C^{(6)} v^2/\Lambda^2$.

\subsection{The $V^{*} \gamma Z$  vertices }
\label{sec:ggZZgZ}

We now apply Eq.~\eqref{eq:mastervertex} to find the form-factors 
$h_3^V$~\cite{Hagiwara:1986vm,Gounaris:1999kf} in $V^*(q)\gamma(k_1) Z(k_2)$-vertices with $V=\gamma,Z$.
For $\gamma^*\gamma Z$ triangle-vertex we need $w=z=1$, 
and a little computation presents a gauge invariant form\footnote{
Although the term proportional to $q^\mu$ is irrelevant for current collider phenomenology
we explicitly write it in all the following vertices to show explicit invariance under gauge symmetry
(for $V=\gamma$) and Bose-symmetry (for $V=Z$).}
\begin{equation}
    \Delta_{\gamma^*\gamma Z}^{\mu\nu\rho}(k_1,k_2) = \frac{s}{M_Z^2}\left ( \epsilon^{\mu\nu\rho\alpha} k_{1\alpha} + \frac{\epsilon^{\nu\rho\alpha\beta} q^\mu k_{1\alpha} k_{2\beta}}{s} \right )
    \, \sum_f M_Z^2\, [A^{(f)}_{3}(s,m_f)]_{\gamma^*\gamma Z}\;,
    \label{eq:ggZ}
\end{equation}
with the associated  form-factor to be
\begin{align}
   e\, h_3^\gamma(s) &= \sum_f M_Z^2 [A^{(f)}_3(s,m_f)]_{\gamma^*\gamma Z} = \sum_f \frac{[\widehat{a}_\gamma^{(f)}\widehat{a}_\gamma^{(f)}\widehat{b}_Z^{(f)}+ \widehat{a}_\gamma^{(f)}\widehat{a}_\gamma^{(f)}\widetilde{b}_Z^{(f)}] }{\pi^2} 
    \nonumber \\[2mm]
    &\times \int_0^1 d x \int_0^{1-x} dy \left [\frac{- M_Z^2\, x y}{M_Z^2 \, x (x-1) - (s-M_Z^2)\, x y +m_f^2} \right ]\;.
    \label{eq:h3gEFT}
\end{align}
In this triangle there is only one dimension-6 insertion. 
The  anomalous terms, i.e. those that depend on the routing momentum parameters $w$ and $z$, 
played a crucial role in getting a gauge invariant result.  
The vertex vanishes when the external gauge bosons are on-shell. The resulting  $h_3^\gamma(s)$ from Eq.~\eqref{eq:h3gEFT} can be written analytically in the limit where the centre of mass energy is greater 
than the top-quark mass threshold,\footnote{The upper limit on $-s-$ comes from the EFT validity.}
\begin{align}
  h_3^\gamma(4 m_t^2 \ll s \ll \Lambda^2)  &\simeq  \frac{e g_Z}{8 \pi^2} \left ( \frac{M_Z^2}{s} \right )
  \, \left ( \frac{v^2}{\Lambda^2} \right ) \, \left [ \left ( C^{\varphi \ell(1)} + C^{\varphi \ell(3)} - C^{\varphi e} \right ) \right. \nonumber \\ 
   &\left. +  \frac{1}{3}\left ( 5\, C^{\varphi q(1)} - 3\, C^{\varphi q(3)} -4\, C^{\varphi u} -  C^{\varphi d} \right )  \right ]
  \nonumber \\[2mm]
  &- i \frac{1}{3\pi} (e g_Z) \left (\frac{M_Z^2}{s}\right ) \left (\frac{m_t^2}{s} \right ) \ln \left (
  \frac{s}{m_t^2} \right)\, \left [ 1 + \frac{v^2}{\Lambda^2} \left ( C^{\varphi q(3)} -
  C^{\varphi q (1)} + C^{\varphi u} \right ) \right ]
  \;. \label{eq:h3glarges}
\end{align}
In this limit, the leading SM effect vanishes due to anomaly cancellation in
Eq.~\eqref{eq:afree}, with the remaining pieces, not shown in \eqref{eq:h3glarges},
descending first as $\ln(s)/s^2$. The imaginary part of $h_3^\gamma$ behaves like $M_Z^2 m_f^2 \ln(s/m_f^2)/s^2$
and therefore it is non-negligible only for the top-quark mass.
Therefore, in the SM EFT, Eq.~\eqref{eq:h3glarges} shows a kind of
“delayed unitarity” at large-$s$.
On the other hand, for energies below the top-threshold, there is an incomplete anomaly cancellation in the SM,
and the SM EFT adds to this vertex, terms proportional to $C v^2/\Lambda^2$:
\begin{eqnarray}
    h_3^\gamma(M_Z^2 \ll  s \ll 4 m_t^2) &\simeq & \frac{e g_Z}{8 \pi^2} \biggl \{
    \left (\frac{M_Z^2}{s} \right )\biggl [\frac{4}{3} + 
    \frac{v^2}{\Lambda^2} \left ( C^{\varphi \ell(1)} + C^{\varphi \ell(3)} - C^{\varphi e} \right ) 
    \nonumber \\[2mm]
    &+& \frac{v^2}{3\,\Lambda^2} \left ( C^{\varphi q(1)} + C^{\varphi q(3)} - C^{\varphi d} \right )\biggr ]  \nonumber \\[2mm]
     &+& \left (\frac{M_Z^2}{9\, m_t^2} \right) \biggl [ 1 + \frac{v^2}{\Lambda^2}\left (C^{\varphi q(3)} - C^{\varphi q(1)} + C^{\varphi u} \right ) \biggr ] \biggr \}\;.\label{eq:h3l}
\end{eqnarray}
By examining Eqs.~\eqref{eq:h3glarges} and~\eqref{eq:h3l}, we see that dimension-6 SM EFT effects 
are evident both in the low- and high-energy regions. Parametrically, however, the high-energy behaviour
of $h_3^\gamma(s)$ in Eq.~\eqref{eq:h3glarges} results in stronger deviations w.r.t the SM because 
the latter scales at most like $\ln(s)/s^2$ due to the chiral anomaly cancellation condition \eqref{eq:afree}.

Similarly, for $Z^*(q)\gamma(k_1)Z(k_2)$-triangle vertex.  Now absence of anomalies with dimension-6 insertions
in $\widehat{Z}\widehat{\gamma}\widetilde{Z}$ requires $w=1, z=1$   while
the triangle $\widetilde{Z}\widehat{\gamma}\widehat{Z}$ needs $w=1, z=-1$ in Eq.~\eqref{eq:mastervertex}. 
For two dimension-6 insertions,
$\widetilde{Z}\widehat{\gamma}\widetilde{Z}$
we need $w=1, \forall z$. Therefore, in this case of a double insertion,  
$z$ is not defined by anomaly cancellation prescription. 
Nevertheless, Bose symmetry defines $z$ precisely. From Table~\ref{tab:1}, Bose symmetry in interchanging 
$i\leftrightarrow k$-legs   with 
$w=1$ results in $z=0$. Matching all this up with \eqref{eq:afree}, 
the anomaly terms in Eq.~\eqref{eq:mastervertex},
give $\sum_f \frac{1}{4\pi^2}\mathcal{A}^{(f)}_{Z\gamma Z} \epsilon^{\mu\nu\rho\alpha} k_{1\alpha}$
for all EFT insertions.
Combining this result with the other terms in Eq.~\eqref{eq:mastervertex}, we find\footnote{We suppress the 
arguments of  $A_i$ and $I_i$-functions in the following paragraphs, i.e., 
$A^{(f)}_{3}=A^{(f)}_{3}(k_1,k_2)$, etc.}
\begin{equation}
    \Delta^{\mu\nu\rho}_{Z^*\gamma Z}(k_1,k_2) = 
    \sum_f \left \{ \frac{1}{2} \left [s A^{(f)}_3 + M_Z^2\, (A^{(f)}_3 + A^{(f)}_5) \right ]\epsilon^{\mu\nu\rho\alpha} k_{1\alpha} + A^{(f)}_3 q^\mu \epsilon^{\nu\rho\alpha\beta} k_{1\alpha} k_{2\beta} \right \}_{Z^*\gamma Z} \;.
\end{equation}
One can show, that the expression in the squared parenthesis is proportional to $(s-M_Z^2)$ for energies 
nearby the $Z$-pole mass and therefore the whole vertex vanishes when all external particles are on-shell. 
It is, therefore, customary to extract the factor $(s-M_V^2)$ and define,
\begin{equation}
   e\, h_3^Z(s) = \frac{M_Z^2}{2(s - M_Z^2)} \sum_f \left [s A^{(f)}_3 + M_Z^2\, (A^{(f)}_3 + A^{(f)}_5) \right ]_{Z^*\gamma Z} \;. \label{eq:h3Zs}
\end{equation}
At low and high-$s$-regime  similar expressions to Eqs.~\eqref{eq:h3glarges} and \eqref{eq:h3l} can be 
derived. In the high-energy region, for example, we find at leading order in $s$,
\begin{align}
    e\, h_3^Z(4 m_t^2 \ll s \ll \Lambda^2) \ \simeq \ \frac{M_Z^2}{2\pi^2 s}\: \sum_f [a_Z^{(f)} a_\gamma^{(f)} b_Z^{(f)}] \ + \ i \frac{3 M_Z^2}{\pi s^2} \:
    [a_Z^{(t)} \:a_\gamma^{(t)} b_Z^{(t)}] \: m_t^2 \ln \left ( \frac{s}{m_t^2} \right )\;,
    \label{eq:h3Zh}
\end{align}
whereas in the low-$s$ region,
\begin{align}
 e\, h_3^Z(M_Z^2 \ll s \ll 4 m_t^2) \ &\simeq \  \frac{M_Z^2}{2 \pi^2 s} \sum_{f\ne t} [a_Z^{(f)} a_\gamma^{(f)} b_Z^{(f)} ] \ - \ \frac{1}{8\pi^2} \left ( \frac{M_Z^2}{m_t^2} \right ) \, [a_Z^{(t)} a_\gamma^{(t)} b_Z^{(t)} ]
 \nonumber \\ &+  \mathcal{O}(M_Z^4/s^2, M_Z^2 s/m_t^4)\;.
 \label{eq:h3Zl}
 \end{align}
By using the results from the Appendix~\ref{app:A} 
for $a^{(f)}_\gamma,a_Z^{(f)}, b_Z^{(f)}$
in Eqs.~\eqref{eq:h3Zh} and
\eqref{eq:h3Zl}  we arrive at  the
dominant contributions to $h_3^Z(s)$ with Wilson coefficients written explicitly 
in analogy with Eqs.~\eqref{eq:h3glarges} and \eqref{eq:h3l} for $h_3^\gamma(s)$.

\subsection{The $V^*ZZ$ vertices}
\label{sec:VZZ}

We  first start with $V=\gamma$. Bose symmetry and  anomaly cancellation following 
the routing momenta described in Table~\ref{tab:1} and paragraph~\ref{sec:Bose}, results
in $\frac{\mathcal{A}_{\gamma ZZ}^{(f)}}{4\pi^2}\epsilon^{\mu\nu\rho\alpha}(k_1-k_2)_\alpha$ 
anomaly term in \eqref{eq:mastervertex}
for both single or double dimesion-6 insertions, which finally, after a little algebra, yields
\begin{equation}
    \Delta_{\gamma^*Z Z}^{\mu\nu\rho}(k_1,k_2) = \frac{s}{M_Z^2}\left [ \epsilon^{\mu\nu\rho\alpha} (k_{1} -k_2)_\alpha + 2 \frac{\epsilon^{\nu\rho\alpha\beta} q^\mu k_{1\alpha} k_{2\beta}}{s} \right ]
    \, \sum_f \frac{1}{2} M_Z^2\, [A^{(f)}_{3}(s,m_f)]_{\gamma^* Z Z}\;.
    \label{eq:gZZ}
\end{equation}
This is a gauge invariant result that vanishes for all particles on-shell as it should. According to the notation (up to a possible overall sign) of Refs.~\cite{Hagiwara:1986vm, Gounaris:1999kf} we find the CP-even form factor 
\begin{align}
e\, f_5^\gamma (s) &=  \sum_f \frac{1}{2} M_Z^2\, [A^{(f)}_{3}(s,m_f)]_{\gamma^* Z Z} 
\nonumber \\[2mm]
&=  \sum_f \frac{[\widehat{a}_\gamma^{(f)} {a}_Z^{(f)} {b}_Z^{(f)}] }{\pi^2} 
    \: \int_0^1 d x \int_0^{1-x} dy \left [\frac{- M_Z^2\, x y}{M_Z^2 \, x (x-1) + M_Z^2 \, y (y-1) - (s-2 M_Z^2)\, x y + m_f^2} \right ]\;,
    \label{eq:f5gEFT}
\end{align}
where, as always denoted in this study, uncharted $a_{V}^{(f)}$ and $b_{V}^{(f)}$ are the full vector and axial-vector couplings in the SM EFT (see Appendix~\ref{app:A}),
or in fact, in any theory extending the Standard Model with a single flavour fermion of mass $m_f$
circulating the triangle. The double integral in Eq.~\eqref{eq:f5gEFT} can be expanded for small or 
large-$s$ and written in a form analogous to Eqs.~\eqref{eq:h3glarges} and~\eqref{eq:h3l}.

The $Z^*ZZ$-vertex can be similarly derived from Eq.~\eqref{eq:mastervertex}. 
For this vertex, also the neutrino is involved directly in the triangle graph. 
However, the analytical form is slightly more complicated here and takes the form
\begin{eqnarray}
    \Delta^{\mu\nu\rho}_{Z^*ZZ}(k_1,k_2) &=& \epsilon^{\mu\nu\rho\alpha}(k_1-k_2)_\alpha
    \sum_f \biggl [ M_Z^2\, (A_3^{(f)} - A_4^{(f)})_{Z^*ZZ} 
    \ + \ \frac{m_f^2}{\pi^2} \, b_Z^{(f)}\, I_{1\, ZZ}^{(f)} \ + \
    \frac{\mathcal{A}_{ZZZ}^{(f)}}{6\pi^2} \biggr ] \nonumber \\[2mm]
    &+& [A_3^{(f)}]_{Z^*ZZ} \, q^\mu \epsilon^{\nu\rho\alpha\beta} 
    k_{1\alpha} k_{2\beta}\;. \label{eq:ZZZ}
\end{eqnarray}
This result vanishes for all $Z$-bosons on-shell \textit{i.e.,} for $s=M_Z^2$, as one can show after a tedious calculation. It is again important to notice that the “anomaly” term proportional to, $\mathcal{A}^{(f)}_{ZZZ}$, is a result of the specific routing of $w$ and $z$ depicted in Fig.~\ref{fig:2} and Table~\ref{tab:1}
for chiral anomaly cancellation and Bose symmetry, respectively. If we had adopted a different routing of momenta,
that is a different prefactor in the anomaly term, Eq.~\eqref{eq:ZZZ} would not vanish on-shell and
would not be Bose symmetric.
In fact, in the SM EFT and at large-$s\ll \Lambda^2$, 
this is the dominant piece in $Z^*ZZ$-vertex form factor $f_5^Z(s)$,
\begin{equation}
   e\, f_5^Z(4 m_t^2\ll s \ll \Lambda^2) \simeq \frac{M_Z^2}{6 \pi^2 s} \sum_f \mathcal{A}^{(f)}_{ZZZ} =
    \frac{M_Z^2}{6 \pi^2 s} \sum_f \biggl [(b_Z^{(f)})^3 + 3\: (a_Z^{(f)})^2\, (b_Z^{(f)}) \biggr ]\;,
\end{equation}
where $a_Z^{(f)}$ and $b_Z^{(f)}$ are the full SM EFT vertices read from Appendix~\ref{app:A}.  
Although the SM contributions 
[for both real and imaginary parts of $f_5^Z(s)$]
vanish as $M_Z^2 m_t^2/s^2$, the SM EFT higher-dimensional pieces do not, and scale like $1/s$. 
As we will see in the next section, this reflects the 
non-decoupling\footnote{Here the terminology “non-decoupling'' simply 
refers to the observable scaling behaviour that survives in the low-energy EFT, 
not to a breakdown of the EFT expansion.}
of the new physics fermions that reside at energy scales of order $\Lambda$. 
This situation, however, can be seen even at low energies far below the top quark threshold,
assuming the latter to be very heavy. Because of the identity,
$m_t^2 \lim_{m_t^2\to \infty} I_1 = -\frac{1}{6} \mathcal{A}_{ZZZ}^{(t)}$, the second term in the
square bracket of Eq.~\eqref{eq:ZZZ} cancels against the 
anomaly term in Eq.~\eqref{eq:ZZZ} leaving behind an incomplete contribution from 
light quarks and leptons
\begin{equation}
 e\, f_5^Z(M_Z^2 \ll s \ll 4 m_t^2) \simeq  \frac{M_Z^2}{6 \pi^2 s} \sum_{f\ne t} 
  \mathcal{A}^{(f)}_{ZZZ} \ - \frac{\mathcal{A}^{(t)}_{ZZZ}}{3\pi^2} \left ( \frac{M_Z^2}{s} \right )
  \left ( \frac{M_Z^2}{4 m_t^2} \right ) + \mathcal{O}(M_Z^2/4 m_t^2)\;,
\end{equation}
that scale like $1/s$ in the SM, affected trivially by SM EFT corrections 
to $Zff$-vertices. The same happens for all vertices (or form factors), 
see for instance Eq.~\eqref{eq:h3l}.

\subsection{nTGVs through four-fermion vertices}
\label{sec:4fermionVs}

Up to now, we analyzed SM EFT effects from the set of operators in \eqref{eq:dim6ops}.
These operators are induced, for example, after the decoupling of heavy fermions,
in a way we shall illustrate in the next section. 
There are, however, contributions to $C$-odd and $P$-odd
form factors $h_3^V(s)$ and $f_5^V(s)$ induced by contact four-fermion vertices. 
Consider, for example, the operator (for simplicity, we assume one flavour generation)
\begin{equation}
    O_{eu} = (\bar{e}_R \gamma_\mu e_R)\, (\bar{u}_R \gamma^\mu u_R)\;,
    \label{eq:eu}
\end{equation}
which  arises from integrating out in an anomaly free UV-theory a heavy scalar leptoquark or a heavy vector boson (see for instance Refs.~\cite{deBlas:2017xtg,Dedes:2021abc}) with appropriate gauge charges and couplings.
We will consider again the amplitude, $e^+(p_1)\, e^-(p_2) \to  V_j (k_1)\, V_k(k_2)$ as a working example. 
The relevant diagram is 
\bigskip
\begin{center}
\begin{fmffile}{4f}
  \parbox{40mm}{\begin{fmfgraph*}(60,51)
    \fmfstraight
    \fmfleft{i1,i2}
    \fmfright{o1,o2}
    \fmf{fermion}{i1,v1}
    \fmf{fermion}{v1,i2}
    \fmf{fermion,tension=0.3}{v1,v2}
    \fmf{fermion,tension=0.3}{v2,v3}
    \fmf{fermion,tension=0.3,label=$u$}{v3,v1}
    \fmf{photon}{v2,o2}
    \fmf{photon}{v3,o1}
\fmfv{decor.shape=circle,decor.filled=full,decor.size=4,f=(1,,.1,,.1)}{v1}
    \fmflabel{$\hspace{-12pt}\nu$}{v2}
    \fmflabel{$\hspace{-12pt}\rho$}{v3}
    \fmflabel{$V_k$}{o1}
        \fmflabel{$V_j$}{o2}
        \fmflabel{$e^-$}{i1}
        \fmflabel{$e^+$}{i2}
            \marrow{ebb}{right}{bot}{$k_1$}{v2,o2}
     \marrow{ecc}{right}{top}{$k_2$}{v3,o1}
\end{fmfgraph*}}
\quad + \quad {\large crossed diagram\;,}
\end{fmffile}
\end{center}
\bigskip
where the red dot is the operator $O_{eu}$ insertion.
For convenience, the rest of  the notation can be
adopted from the diagram above Eq.~\eqref{eq:trigono} and Fig.~\ref{fig:1}.
This triangle diagram is linearly divergent, and similar techniques to those discussed previously apply.
The WIs \eqref{eq:WI2} and \eqref{eq:WI3} in legs $V_j$ and $V_k$ must be respectively
satisfied. Consequently, 
there is an  additional contribution to  the amplitude 
$e^+(p_1) \, e^-(p_2) \to V^\nu_j(k_1)\, V^\rho_k(k_2)$  of Eq.~\eqref{eq:amp}, which reads,
\begin{equation}
    - \frac{1}{2}\, [\bar{v}(p_1)\, \gamma_\mu (1+\gamma^5) \, u(p_2)]\, \Delta_{\widetilde{V}_iV_jV_k}^{\mu\nu\rho} (k_1,k_2;w,z)\;.
    \label{eq:4fD}
\end{equation}
The vertex $\Delta_{\widetilde{V}_iV_jV_k}^{\mu\nu\rho}$ is given in Eq.~\eqref{eq:mastervertex}.
The arbitrary parameters $w$ and $z$ are defined by satisfying the WIs. These are $w=1$ and $z=-1$ for only one 
SM EFT insertion in vertex-$i$. In this case, the fictitious gauge boson $\widetilde{V}_i$ has
only SM EFT couplings to fermions. For example, for the operator \eqref{eq:eu}, we have~\cite{dedes:2017zog}
\begin{equation}
    a_{\widetilde{V}_i}^{(eu)} = b_{\widetilde{V}_i}^{(eu)} = \frac{C^{eu}}{2\Lambda^2} \;.
\end{equation}
For example, the contribution \eqref{eq:4fD} emerges at low energies after decoupling a 
heavy gauge boson (such as $Z'$) from a UV-theory .

\section{Top-down approach for chiral anomalies and nTGVs}
\label{sec:topdown}

An examination of a specific UV-model can enhance and clarify the bottom-up approach 
to chiral anomalies and TGVs, discussed in the preceding sections.
We will focus on the chiral anomalies and decoupling effects 
of an anomaly-free renormalizable gauge theory, which adds to the SM a heavy vector-like electron $E$.
We will next show how anomalies cancel in the full theory of this UV-model after electroweak symmetry breaking.
We then integrate out the heavy particles and derive  the SM EFT of this model,
verifying the bottom-up picture we found in section~\ref{sec:AnomSMEFT}.
As an illustration, we also calculate the nTGV form-factor $h_3^\gamma(s)$ and compare it with our 
SM EFT results in section~\ref{sec:nTGC}.
As far as the decoupling of heavy fermions is concerned, this simple model incorporates 
most of the necessary ingredients for studying the chiral anomalies and nTGVs in a top-down approach.

\subsection{A simple UV-theory example}
\label{sec:fulltheory}
Consider a model with a heavy vector-like electron $E(1,1,-1)$ 
which is a singlet under $SU(3)_c$ and $SU(2)_L$ gauge groups
but has hypercharge identical to the right-handed electron, $Y_E=-1$. The field $E$ has a Dirac mass $M$
which is assumed much heavier than the electroweak scale, $M\gg v$, with $v$ the vacuum 
expectation value of the Higgs field, $\varphi$. The full Lagrangian is 
\begin{equation}
    \mathcal{L} = \mathcal{L}_{\rm SM} + \mathcal{L}_{\rm BSM}\;, 
\end{equation}
where $\mathcal{L}_{\rm SM}$ is the SM Lagrangian\footnote{We use the notation of Ref.~\cite{dedes:2017zog}
throughout.} and 
\begin{equation}
    \mathcal{L}_{\rm BSM} = \bar{E}_L i \slashed{D} E_L +  \bar{E}_R i \slashed{D} E_R 
    -M (\bar{E}_L E_R + \bar{E}_R E_L ) - y_E\: \bar{\ell}_L \cdot \varphi\, E_R - 
    y_E^*\: \bar{E}_R\, \varphi^\dagger \cdot \ell_L \;,
    \label{eq:LBSM}
\end{equation}
where $D_\mu = \partial_\mu + i g' Y_E B_\mu$ is the covariant derivative and $B_\mu$ the
gauge field associated with the hypercharge symmetry.\footnote{In $\mathcal{L}_{\rm BSM}$ 
terms like $m \bar{E}_L e_R +\mathrm{h.c.}$, that mix heavy and light electron fields, are also allowed. It is 
easy to show that these terms can be eliminated by a redefinition of $E_R$ and $e_R$ fields, leaving 
\eqref{eq:LBSM} as a starting BSM Lagrangian.} The full Lagrangian $\mathcal{L}$ is gauge anomaly free,
since we have added to the SM-Lagrangian a Dirac fermion $E$, which in turn means setting the vector-like condition, $b^{(E)}_Z=0$, 
in the anomaly factor Eq.~\eqref{eq:anterms}. The axial current, $J_A^\mu= \bar{E}\gamma^\mu \gamma^5 E$, associated with the 
chiral symmetry, $E\to e^{i\theta \gamma^5} E$, 
is never conserved, $\partial_\mu J_A^\mu = 2 i M \bar{E} \gamma^5 E$ because  
$M$ is the mass of the “heavy” vector-like fermion and therefore, the limit $M\to 0$ is unattainable by definition.

In broken EW phase, $SU(2)_L\times U(1)_Y\to U(1)_{\rm EM}$, 
the heavy $(E)$ and light $(e)$ electron fields' mass terms mix each other
through the vev $v$ of the Higgs field 
$\varphi^T=(G^+, \varphi_0)$ 
with $\varphi_0=\frac{1}{\sqrt{2}}(v+h+i G^0)$ and the Yukawa coupling $y_E$. By 
assuming that the electron Yukawa coupling, $y_e \bar{\ell}_L \varphi e_R + \mathrm{h.c}$, 
is zero for simplicity, we obtain 
\begin{equation}
    \mathcal{L}^{\rm mass}_{\rm BSM} = - \left (\begin{array}{cc}
       \bar{e}_L  & \bar{E}_L \end{array} \right )
      \, \mathcal{M}\,  \left (\begin{array}{c} e_R \\ E_R \end{array} \right) - \left (\begin{array}{cc}
       \bar{e}_R  & \bar{E}_R \end{array} \right )
      \, \mathcal{M}^\dagger \,  \left (\begin{array}{c} e_L \\ E_L \end{array} \right)\;, 
      \label{eq:Lm}
\end{equation}
with the mass matrix $\mathcal{M}$ being 
\begin{equation}
    \mathcal{M} = \left ( \begin{array}{cc}
       0  &  y_E \frac{v}{\sqrt{2}} \\[2mm]
       0  & M
    \end{array} \right )\;.
\end{equation}
Applying the singular value decomposition theorem~\cite{0521386322} we obtain $\mathcal{M} = U_e D_e W_e^\dagger$ where $D_e$ the diagonal mass matrix 
\begin{equation}
    D_e = \left ( \begin{array}{cc}
        0 & 0 \\
        0 & M \, \sqrt{1 + |\epsilon|^2 }
        \end{array} \right ) \;, 
        \qquad \mathrm{with} \qquad \epsilon \equiv \frac{y_E v}{\sqrt{2} M} \;.
        \label{eq:De}
\end{equation}
%
The matrices $U_e$ and $W_e$ are unitary, diagonalizing the Hermitian 
 matrices $\mathcal{M}\mathcal{M}^\dagger$ and 
 $\mathcal{M}^\dagger\mathcal{M}$, respectively.
Explicitly, they are written as
%
\begin{equation}
    U_e = \frac{1}{\sqrt{1+|\epsilon|^2}} \, \left ( \begin{array}{cc}
      1   &  \epsilon \\
      -\epsilon^*   & 1 
      \end{array} \right ) \;, \qquad
    W_e = \left ( \begin{array}{cc}
        1 & 0 \\
        0 & 1 
        \end{array} \right ) \;.
        \label{eq:UeWe}
\end{equation}
There are few remarks to be stated here: First, the mass of the heavy fermion $E$ is shifted upwards 
from $M$ to $M'=M \sqrt{1+|\epsilon|^2}$ in the new (mass) basis,
while the light electron stays massless (since we have set $y_e=0$); by increasing $y_E$ the
scale $M'$ will increase as well for fixed $M$.
Second, the transformation to the new basis affects \textit{only} the left-handed, heavy and light, electrons
\begin{equation}
    \left ( \begin{array}{c}
          e_L'  \\ E_L'
     \end{array} \right ) = U_e^\dagger \, \left ( \begin{array}{c}
          e_L  \\ E_L
     \end{array} \right )\;,
\end{equation}
but not the right-handed ones [see Eq.~\eqref{eq:UeWe}]; in other words, it is potentially anomalous.
Although in the new “prime” basis, we have decoupled propagators between heavy $(E')$ and light fermions $(e')$,
\begin{equation}
 \mathcal{L}^{\rm mass}_{\rm BSM} = \left ( \begin{array}{cc}
     \bar{e}'   &  \bar{E}' 
     \end{array} \right ) D_e \left ( \begin{array}{c}
          e'  \\
          E' 
     \end{array} \right )\;,
\end{equation}
contributions to the triangle anomalies will show up in the interaction of gauge boson vertices with
heavy-light electrons. It is straightforward to find the relevant interaction terms with 
the currents associated to the weak gauge bosons
$W^\pm,Z$ and the photon $A$,
\begin{align}
    \mathcal{L} &\supset -g \, (W_\mu^+ J_W^{\mu\, +} + W_\mu^- J_W^{\mu\, -}) - g_Z Z_\mu J_Z^\mu - e A_\mu J_{\rm EM}^\mu \;, \label{eq:Lgauge}\\[2mm]
    J_W^{\mu\, +} &= \frac{1}{\sqrt{2}} \left [ \bar{\nu}_L \gamma^\mu (U_e)_{11} e_L' + 
    \bar{\nu}_L \gamma^\mu (U_e)_{12} E_L' \right ]\;, \qquad J_W^{\mu\, -} = (J_W^{\mu \, +})^\dagger \;,
    \label{eq:Jw}\\[2mm]
    J_Z^\mu &= \bar{\nu}_L\gamma^\mu \left (\frac{1}{2} \right ) \nu_L + 
    \bar{e}_L'\gamma^\mu \left(-\frac{1}{2} |(U_e)_{11}|^2 +s_w^2 \right ) e_L' + 
    \bar{E}_L'\gamma^\mu \left(-\frac{1}{2} |(U_e)_{12}|^2 +s_w^2 \right ) E_L' \nonumber \\
    &+ \bar{e}_L' \gamma^\mu \left [-\frac{1}{2} (U_e)_{11}^* (U_e)_{12} \right ] E_L'
    + \bar{E}_L' \gamma^\mu \left [-\frac{1}{2} (U_e)_{12}^* (U_e)_{11} \right ] e_L' \nonumber \\
    &+ \bar{e}_R' \gamma^\mu (s_w^2) e_R' +  \bar{E}_R' \gamma^\mu (s_w^2) E_R' \label{eq:Jz}\\[2mm]
    J^\mu_{\rm EM} &= \bar{e}_L' \gamma^\mu (-1) e_L' + \bar{e}_R' \gamma^\mu (-1) e_R'
    + \bar{E}_L' \gamma^\mu (-1) E_L' + \bar{E}_R' \gamma^\mu (-1) E_R' \;, \label{eq:Jem}
\end{align}
with $g$ being the $SU(2)_L$-gauge coupling, $g_Z=\sqrt{g'^2+g^2}=e/s_W c_W$,\footnote{It is $s_W \equiv \sin(\theta_W)$ and 
$c_w\equiv\cos(\theta_w)$ with $\theta_w$ being the weak mixing angle.} and $e$ the electron electric charge.
Obviously, from Eqs.~\eqref{eq:Jw} and~\eqref{eq:Jz}, there are heavy-light fermion mixed vertices in the 
weak charged and neutral currents but not for the electromagnetic current, Eq.~\eqref{eq:Jem}. 
As we demonstrate below, chiral anomalies cancel out if and only if we consider these interactions.
Turning all fermions into Dirac spinors and reading vector ($a_V^{(f)}$) and vector-axial couplings
($b_V^{(f)}$) from the vertex 
\vspace{0.3cm}
\begin{center}
\begin{fmffile}{Z0}
  \parbox{20mm}{\begin{fmfgraph*}(40,40)
    \fmfstraight
    \fmfleft{i1}
    \fmfright{o1,o2}
    \fmf{photon}{i1,v1}
    \fmf{fermion,tension=0.3}{o1,v1}
    \fmf{fermion,tension=0.3}{v1,o2}
        \fmflabel{$V$}{i1}
    \fmflabel{$f$}{o1}
        \fmflabel{$f$}{o2}
\end{fmfgraph*}}
{$ \ = \ -i \, \bar{f}\, \gamma^\mu (a_V^{(f)} + b_V^{(f)} \gamma^5)\, f \;, \label{dd} $ }
\end{fmffile}
\end{center}
\begin{eqnarray}
    \label{fd:Zff}
\end{eqnarray}
\vspace{0.3cm}
with $V=Z,\gamma$ we find
\begin{align}
   & a^{(\nu)}_Z = \frac{g_Z}{4}\;, \quad b^{(\nu)}_Z = -\frac{g_Z}{4}\;, \quad a^{(\nu)}_\gamma=0 \;, \\
   & a^{(e')}_Z = \frac{g_Z}{2} \left (-\frac{1}{2} |(U_e)_{11}|^2 + 2 s_w^2 \right )\;, \quad 
   b^{(e')}_Z = \frac{g_Z}{4} |(U_e)_{11}|^2 \;, \quad a^{(e')}_\gamma = -e \;,   \label{eq:eeZ}\\
   & a^{(E')}_Z= \frac{g_Z}{2} \left (-\frac{1}{2} |(U_e)_{12}|^2 + 2 s_w^2 \right )\;, \quad 
   b^{(E')}_Z = \frac{g_Z}{4} |(U_e)_{12}|^2 \;, \quad a^{(E')}_\gamma = -e \;,\label{eq:ZEb}\\
   & a^{(e'-E')}_Z= -\frac{g_Z}{4} (U_e)_{11}^* (U_e)_{12}\;, \quad 
     b^{(e'-E')}_Z = \frac{g_Z}{4} (U_e)_{11}^* (U_e)_{12} \;, \quad a^{(e'-E')}_\gamma = 0 \;,
\end{align}
plus $a_V^{(q)}$ and $b_V^{(q)}$ for quarks as in the SM.\footnote{In this model with heavy vector-like electrons,
the quark sector remains untouched w.r.t the SM. Recall, however, that the pairs $a_V^{(q)}$ and $b_V^{(q)}$
for $q=u,d$ are necessary to cancel the chiral anomalies. In our notation, they can be read directly from  Eq.~(26)
of Ref.~\cite{Dedes:2012me}.} Also from Eq.~\eqref{eq:Jz}, it is $a^{(E'-e')}_Z=[a^{(e'-E')}_Z]^*$
and $b^{(E'-e')}_Z=[b^{(e'-E')}_Z]^*$. Of course, there are no axial-vector couplings to the photon, $b_\gamma^{(f)}=0$.

To check the WIs in Eqs.~\eqref{eq:WI1}\eqref{eq:WI2} and \eqref{eq:WI3} we also need the neutral Goldstone
couplings to fermions. They can easily be found to be\footnote{There are no $G^0 \bar{e}'\gamma^5 e'$
terms because we have set $y_e$ to zero for simplicity.}
\begin{align}
    \mathcal{L}^{G^0} = &-(i G^0 \bar{E}' \gamma^5 E')\, \left ( \frac{M'}{v} |(U_e)_{12}|^2 \right )
    -(i G^0 \bar{e}' P_R E')\, \left ( \frac{M'}{v} (U_e)_{11}^* (U_e)_{12} \right ) \nonumber \\
    &+(i G^0 \bar{E}' P_L e')\, \left ( \frac{M'}{v} (U_e)_{11} (U_e)_{12}^* \right )\;.
    \label{eq:G0EE}
\end{align}
In pure analogy with vector boson vertices to fermions, here again, there are contributions
to Goldstone-triangles from heavy and heavy/light fermion vertices. Moreover, and not without a reason,
these couplings look alike to the corresponding $V\bar{f}f$ couplings in Eq.~\eqref{eq:Jz}.
They are made as such to satisfy the WI in Eq.~\eqref{eq:WIgeneral}.

Even at this point, we can contact the EFT expansion of the theory. The true 
expansion parameter here is $\epsilon$ defined in Eq.~\eqref{eq:De}. Then
from Eq.~\eqref{eq:UeWe} we can expand the $U_e$-matrix elements and write
\begin{equation}
    |(U_e)_{11}|^2 = 1 - |\epsilon|^2 + |\epsilon|^4 + [\mathcal{O}(\epsilon)]^6\;, 
    \qquad 
    |(U_e)_{12}|^2 = |\epsilon|^2 - |\epsilon|^4 + [\mathcal{O}(\epsilon)]^6\;.
    \label{eq:Ueexpansion}
\end{equation}
It is interesting that we can directly read the Wilson coefficients in SM EFT arising 
from the decoupling of the heavy electron $E'$ 
by matching the Feynman Rule for the three-point $Zee$ vertex
of the model at hand [see Eq.~\eqref{eq:eeZ}]
with the corresponding SM EFT Feynman Rule in Appendix~\ref{app:A} in broken phase.
By simultaneously noting that, 
for this particular model, the $Z\nu\nu$-vertex is the same as in the SM, we find easily
the matching condition,
\begin{equation}
    C^{\varphi e} = 0\;,\qquad C^{\varphi \ell (1)} = C^{\varphi \ell (3)} = -\frac{|y_E|^2}{4}\;,
    \qquad \Lambda = M\;,
    \label{eq:Cphil}
\end{equation}
in agreement with Ref.~\cite{deBlas:2017xtg}.
With this in mind, we now possess all the necessary components to ascertain if chiral anomaly cancellation exists and we convey a few representative nTGVs for this particular model.
We show these with the following examples.

\subsubsection{Example: Anomaly cancellation in $Z\gamma\gamma$ triangle and a nTGV}
\label{sec:Zgg}

This is the simplest situation with only one possible AVV-type anomaly. 
The anomaly factor of Eq.~\eqref{eq:anterms} for light and heavy 
electrons contains,
\begin{align}
   \sum_{f=e',E'} \mathcal{A}^{(f)}_{Z\gamma\gamma} &= b_Z^{(e')} a_\gamma^{(e')} a_\gamma^{(e')} + b_Z^{(E')} a_\gamma^{(E')} a_\gamma^{(E')} \nonumber \\
   &= \frac{g_Z e^2}{4} \left [|(U_e)_{11}|^2 + |(U_e)_{12}|^2 \right ] = \frac{g_Z e^2}{4}\;, \label{eq:ZAA}
\end{align}
where the last step  follows from the fact that $U_e$ is a unitary matrix. Needless to say,
the anomaly piece of Eq.~\eqref{eq:ZAA} cancels when we add the contribution from up and down quarks.
In this UV-complete theory, 
the cancellation happens independent of the routing of the momenta. However, failure to include the 
heavy fermions $E'$ in \eqref{eq:ZAA}, 
even if those are vector-like in the unbroken phase, will cause a chiral anomaly in the broken phase of 
order $|y_E|^2 v^2/M^2$. As we will show below, the effect of the heavy particle $E'$ triangle loop is magically
implemented in the SM EFT through the Goldstone boson vertex to SM fermions.

One could also check that pieces from finite anomalous terms arising 
from Goldstone ($G^0\gamma\gamma)$ and gauge-boson ($Z\gamma\gamma$) triangle diagrams
cancel out explicitly by themselves satisfying the WIs in Eqs.~\eqref{eq:WI1}\eqref{eq:WI2} and \eqref{eq:WI3}.
This is almost obvious from Eq.~\eqref{eq:trigono2} with $m_e'=0$ and $m_E'=M'$, and, from Eq.~\eqref{eq:trigonoG}
with the Goldstone and $Z$-boson vertices to heavy electrons $E'$ taken from Eqs.~\eqref{eq:G0EE} and \eqref{eq:ZEb}, respectively.

The anomaly factor in Eq.~\eqref{eq:ZAA} is the same for all interchanges of the 
external legs 
$\mathcal{A}^{(f)}_{Z\gamma\gamma}=\mathcal{A}^{(f)}_{\gamma Z\gamma}=\mathcal{A}^{(f)}_{\gamma\gamma Z}$.
An anomaly free gauge extension of the SM does not mean that there are no effects in the triple gauge
boson vertices, although the truth is that the anomaly cancellation condition, 
$\sum_f \mathcal{A}^{(f)}=0$,   hugely suppresses these effects.
It is  interesting, phenomenologically at least, to find the $\gamma(q)^*\gamma(k_1) Z(k_2)$ vertex where 
the incoming photon is off-shell and the outgoing photon and $Z$-boson are on-shell particles. 

Using the formula \eqref{eq:ggZ}, or instead, directly Eq.~\eqref{eq:h3gEFT} for the model at hand,  we find,
in the “intermediate” energy region between the top threshold and the high scale $M$,
\begin{equation}
    \Re e [h_3^\gamma(s)] \ \simeq \ -\frac{g_Z e}{8 \pi^2} \left (\frac{|y_E|^2 v^2}{M'^{2}} \right ) \left [
    \frac{M_Z^2}{2\, s} + \frac{1}{24}\frac{M_Z^2}{M'^2}+\dots \right ]\;, \quad 4\, m_t^2 \ll s \ll M'^2 \;.
    \label{eq:h3gbelowM}
\end{equation}
The effect on $h_3^\gamma(s)$ from the decoupling of a heavy electron $E$ is twofold: 
first a dimension-6 contribution\footnote{That this is precisely the dimension-6 SM EFT contribution we can see by applying the Wilson-coefficients from \eqref{eq:Cphil}
 to Eq.~\eqref{eq:h3glarges} we derived earlier in section~\ref{sec:ggZZgZ}.} which, however, 
 drops off with the square of the centre of mass energy like
$1/s$ and, second, a constant with energy 
dimension-8 contribution. The “$1/2s$-term”  arises because of the reduced $Ze'e'$-vertex relative to the SM by 
a factor $|(U^e)_{11}|^2$ which does not cancel in the effective SM anomaly cancellation condition. On the
other hand, the “$1/24 M'^2$-term” appears from the decoupling of the heavy $E'$ with a suppression prefactor.

Hence, it appears from Eq.~\eqref{eq:h3gbelowM} that the dimension-6 contribution is \textit{always}   more 
important than the dimension-8 for centre of mass energies $s<M'^2$. 
To our knowledge, this fact has been overlooked in literature,
where sometimes opposite conclusions are stated.
Although the dimension-8 contribution\footnote{Following Ref.~\cite{Cepedello:2024ogz},
the dimension-8 contribution arises from a linear
combination of Wilson coefficients associated with four operators of the kind ---$\varphi^\dagger \widetilde{V}^{\mu\nu} (D^\rho V_{\nu\rho} ) D_\mu \varphi$--- with $(\widetilde{{V}}_{\mu\nu})\; V_{\mu\nu}$ being the 
(dual) field-strength tensor for either $U(1)_Y$ or $SU(2)_L$ gauge group.} of Eq.~\eqref{eq:h3gbelowM}  agrees with recent literature~\cite{Cepedello:2024ogz,Ellis:2024omd},
the dimension-6 term of Eq.~\eqref{eq:h3gbelowM} has been entirely ignored there.

One may think that a conclusion from Eq.~\eqref{eq:h3gbelowM} 
is a result particularly attributed to the considered  model with a heavy vector-like electron.
We believe, however, that is generic to all models of decoupling anomaly free clusters of heavy fermions
with portal Yukawa couplings to the SM (like $y_E$ here).
If the dimension-8 term was larger than the $1/s$ term, then violation of unitarity would take place,\footnote{By 
definition, $h_3^\gamma(s)$ is a monotonically decreasing function away from particle thresholds.}
contrary to our consideration for a renormalizable and anomaly free theory at high energies.

Furthermore, in the low $s$-region below the top-threshold, we find
\begin{eqnarray}
    h_3^\gamma(s)\ \simeq \
    \frac{e g_Z}{8\pi^2} \biggl [ \left ( \frac{M_Z^2}{s} \right ) \left (\frac{4}{3} - 
    \frac{|y_E|^2 v^2}{2 M'^{2}} \right ) + \left (\frac{M_Z^2}{9\, m_t^2} - \frac{|y_E|^2}{24} \frac{v^2 M_Z^2}{M'^{4}} \right ) 
    \biggr ]\;, \quad M_Z^2 \ll s \ll 4 m_t^2\;, \label{eq:h3glfull}
\end{eqnarray}
with a negligible imaginary part that is not shown.
In this case, chiral anomalies do not cancel out due to the top-quark mass decoupling and the 
SM contribution is enhanced due to the light lepton and quark contribution. 
The dimension-six contribution is the same as in Eq.~\eqref{eq:h3gbelowM}, and in this particular model decreases 
the values of $h_3^\gamma$ in full agreement with the SM EFT outcome of Eq.~\eqref{eq:h3l} plugged in with
the matching condition $\eqref{eq:Cphil}$. The dimension-8 term, i.e., the last term
in \eqref{eq:h3glfull}, is completely negligible in this region.

Finally, in the energy region above the heavy fermion $E$-mass $s\gg M'^2$, the form factor 
$h_3^\gamma(s)$ drops off much faster\footnote{In fact as 
$h_3^\gamma(s\gg M'^2) \simeq \frac{M_Z^2 M'^2}{s^2} [ \ln^2 (\frac{s}{M'^2}) -\pi^2]$.}  than $1/s$ due
to the exact chiral anomaly cancellation [see Eq.~\eqref{eq:ZAA} and the discussion below].
We have numerically calculated $h_3^\gamma(s)$ without making approximations 
and discuss all the above results quantitatively in section~\ref{sec:pheno}.

\subsubsection{Example: Anomaly cancellation in $Z\gamma Z$ triangle and a nTGV}

In this case, there are also mixed $Z-E'-e'$-vertices in the triangle. The part of the anomaly factor 
for $e'$ and $E'$ is
\begin{align}
 \sum_{f=e',E'} \mathcal{A}^{(f)}_{Z\gamma Z} \ &= \   \sum_{f=e',E'} \left [b_Z^{(f)} a_\gamma^{(f)} a_Z^{(f)} + a_Z^{(f)} a_\gamma^{(f)} b_Z^{(f)} \right ]
      \nonumber \\[2mm]  & = 2 b_Z^{(e')} a_\gamma^{(e')} a_Z^{(e')} + 2 b_Z^{(E')} a_\gamma^{(E')} a_Z^{(E')} 
    + 2 b_Z^{(E'-e')} a_\gamma^{(e')} a_Z^{(e'-E')} + 2 b_Z^{(e'-E')} a_\gamma^{(E')} a_Z^{(E'-e')}
    \nonumber \\[2mm]
    & = - \frac{g_Z^2 e}{4} \left \{ -\frac{1}{2} \biggl [ |(U_e)_{11}|^2 + |(U_e)_{12}|^2 \biggr ]^2
    + 2 s_w^2 \biggl [ |(U_e)_{11}|^2 + |(U_e)_{12}|^2 \biggr ] \right \}
    \nonumber \\[2mm]
    & = - \frac{g_Z^2 e}{4} \left ( -\frac{1}{2} + 2 s_w^2 \right ) \;.
    \label{eq:ZgZ}
\end{align}
This is precisely the opposite of the anomaly produced by the up and down quarks.
The last step in \eqref{eq:ZgZ} follows from the unitarity of $U_e$-matrix.  
Again, the heavy electrons do not completely decouple: 
apart from the heavy triangle loop, there are additional triangles in \eqref{eq:ZgZ} with
mixed $Z-e'-E'$ vertices of \eqref{eq:Jz} which are essential for the  anomaly cancellation.
Obviously,  the cancellation
of chiral anomalies is an all orders result in the $\epsilon$-expansion, defined in Eq.~\eqref{eq:Ueexpansion}.

Additionally, we have carried out the calculation 
of the form-factor $h_3^Z(s)$ defined in Eq.~\eqref{eq:h3Zs}. At high energies, the real part is
\begin{equation}
    \Re e[h_3^Z(s)] = - \frac{y_E^2}{4 \pi^2}\, \frac{g_Z^2 v^2}{M'^2}\, \biggl [ 
    \frac{(1-2 s_W^2)}{8} \frac{M_Z^2}{s} + \frac{(4-3 s_W^2)}{144}\, \frac{M_Z^2}{M'^2}+\dots\biggr ]\;, \quad 4 m_t^2\ll s\ll M'^2\;.
\end{equation}
The first term in the square brackets agrees with the dimension-6 result of Eq.~\eqref{eq:h3Zh} 
after replacing the matching outcome of Eq.~\eqref{eq:Cphil}. The second term inside the square 
brackets is the dimension-8 contribution to $h_3^Z(s)$. Similar to Eq.~\eqref{eq:h3gbelowM}, this
is \textit{always} more than an order of magnitude smaller than the dimension-6 contribution. In the 
low-energy region below the top-threshold, we find
\begin{align}
    h_3^Z(s) = -\frac{g_Z^2}{4\pi^2} \biggr \{  \frac{M_Z^2}{2s}
    &\biggl [ \biggl (-\frac{1}{2}+\frac{4}{3} s_W^2\biggr ) + \frac{y_E^2 v^2}{M'^2} \frac{(1- 2 s_W^2)}{4}
    \biggr ] \nonumber \\
    + &\biggl [ \frac{M_Z^2}{24 m_t^2} \biggl (-\frac{1}{2}+\frac{4}{3} s_W^2\biggr ) +\frac{(y_E^2 v^2) M_Z^2}{M'^4} \frac{(4-3s_W^2)}{144} \biggr ] + \dots \biggr \}\;, \quad M_Z^2\ll s\ll 4 m_t^2 \;.
\end{align}
In this region, the SM contribution is the dominant one, although there is a tendency to cancel 
the dimension-6 contribution for considerable Yukawa coupling $y_E \simeq 5$ where, however, 
the $\epsilon$-expansion is 
questionable [\textit{c.f.} discussion in section~\ref{sec:pheno}]. 
For $y_E\sim O(1)$ and $M=1~\mathrm{TeV}$, the NP dimension-6 effect from 
the heavy electron is approximately, $5\% \times y_E^2$ regarding the SM expectation. 
The dimension-8 contribution is more than an order of magnitude smaller than that.

\subsubsection{Example: Axial-Axial-Axial (AAA) anomaly cancellation in $Z Z Z$ triangle }

In our final example, we show the cancellation of the AAA-anomaly in this model with heavy 
vector-like electrons. The AAA-anomaly factor when adding $e',E'$ and the neutrino $\nu$ 
in the triangle loop must vanish (as in the SM). From an analytical standpoint, it gives exactly this:
\begin{align}
 &\sum_{f=\nu, e',E'} \mathcal{A}^{(f)}_{ZZZ \mathrm{(axial-part)}} \ = \ 
 \sum_{f=\nu, e',E'} b_Z^{(f)}b_Z^{(f)}b_Z^{(f)}
 \nonumber \\[2mm]
 &= b_Z^{(\nu)}b_Z^{(\nu)}b_Z^{(\nu)} +b_Z^{(e')}b_Z^{(e')}b_Z^{(e')}
 + b_Z^{(E')} b_Z^{(E')} b_Z^{(E')} + 3 b_Z^{(e'-E')}b_Z^{(E'-e')}b_Z^{(e')} + 
 3 b_Z^{(E'-e')} b_Z^{(e'-E')} b_Z^{(E')} \nonumber \\[2mm]
 &= \left ( \frac{g_Z}{4} \right )^3 \left \{ -1 + \biggl [ |(U_e)_{11}|^2 + |(U_e)_{12}|^2 \biggr ]^3 \right \} 
 \nonumber \\[2mm]
 &= 0 \;.
\end{align}
The factor 3 in the RHS of the second line accounts for all possible diagram topologies 
with $e'$ and $E'$ in the triangle. 
The result is as expected from the SM: AAA-chiral anomalies cancel 
among leptons and quarks separately. Again, the cancellation is exact to all orders in $\epsilon$-expansion
of Eq.~\eqref{eq:Ueexpansion} thanks to the unitarity of the $U_e$-matrix.

\subsubsection{Remarks}

The principal points highlighted in the aforementioned examples 
regarding the cancellation of chiral anomalies after electroweak symmetry breaking are:
\begin{itemize}
    \item Addition of  leptons and quarks within a generation.
    \item Inclusion of heavy and light fermions in the triangle loop.
    \item Unitarity of the $U_e$-matrix. 
    \item Anomaly cancellation can also be seen order by order in $|\epsilon|^2 = |y_E|^2 v^2/2 M^2$ expansion.
\end{itemize}
In this linear realization of symmetry breaking, the Goldstone boson vertices to fermions result in finite
contributions in triangle diagrams and have nothing to do with chiral anomalies. 
Neither is how the routing of the momenta in one-loop triangles is defined.

In all the above, we have set the electron Yukawa coupling to zero, $y_e=0$. 
However, the result of the cancellation of the anomalies remains the same for $y_e\ne 0$; just the matrix
$U_e$ will be more detailed, but it will still be a unitary matrix.

\subsection{SM EFT after the decoupling of a heavy vector-like electron}
\label{sec:SMEFT}

The effective action at low energies consists of light SM fields, collectively called $e$ in the following,
is obtained after integrating out from the path integral a heavy fermion field $E$ of Eq.~\eqref{eq:LBSM}~\cite{Weinberg:1980wa,Witten:1976kx,Kazama:1981fx,Bilenky:1993bt}:
\begin{align}
    e^{i S_{\rm eff}[e]} &= \int dE\, d\bar{E} \, e^{i S[e, E]} = \int dE\, d\bar{E} \, e^{i S_{\rm SM}[e]} \, 
    e^{i S_{\rm BSM}[e,E]} \nonumber \\[2mm]
    &= e^{i S_{\rm SM}[e]} \, \int dE\, d\bar{E} \, e^{i \int d^4 x \left \{ \bar{E} (i \slashed{D} ) E
    -M \bar{E} E - y_E \bar{\ell} \cdot \varphi P_R E - y_E^* \bar{E} \varphi^\dagger P_L \cdot \varphi
    \right \} }\nonumber \\[2mm]
    &= e^{i S_{\rm SM}[e]} \, \int dE\, d\bar{E} \, e^{i \int d^4 x \left \{ \bar{E} K E + \bar{J} E
    + \bar{E} J \right \}} \;, \label{eq:pi}
\end{align}
where $K=i\slashed{D} -M$ and $\bar{J}=-y_E \bar{\ell} \cdot \varphi P_R $ and 
$J = - y_E^*  \varphi^\dagger P_L \cdot \varphi$. We may think of $J$ and $\bar{J}$ as “sources
and sinks” for the heavy fields, $\bar{E}$ and $E$, 
because without $J$s the fields $E$ decouple completely. Solving 
the path integral in \eqref{eq:pi} we find
\begin{align}
    e^{i S_{\rm eff}[e]} &= e^{i S_{\rm SM}[e]} \, 
    e^{-i \int d^4 x\, \bar{J}(x)\,  (i \slashed{D} - M)^{-1} \, J(x)}\, 
    \det [i\slashed{D}-M]\;, \label{eq:eiS}
\end{align}
with the covariant derivative now acting on light SM fields. The heavy fields $E$ have been 
integrated out from the path integral. Expanding the operator $(i\slashed{D}-M)^{-1}$
in powers of small momenta ($p\ll M)$, we find
\begin{equation}
    (i\slashed{D}-M)^{-1} \ = \ -\frac{1}{M} \left ( 1 + \frac{i\slashed{D}}{M} + \frac{(i\slashed{D})^2}{M^2} +
    \frac{(i\slashed{D})^3}{M^3} + \dots \right ) \;.
\end{equation}
Only odd powers of $(i\slashed{D})$ survive in the exponent of Eq.~\eqref{eq:eiS}. 
At the zeroth order in $\hbar$-expansion, Eq.~\eqref{eq:eiS} results in the local effective action, $S_{\rm eff}=\int d^4 x \mathcal{L}_{\rm eff}$, which up to dimension-8 operators reads
\begin{align}
    \mathcal{L}_{\rm eff} \ = \ \mathcal{L}_{\rm SM} \ +\  \frac{|y_E|^2}{M^2} (\bar{\ell}\cdot \varphi) (i \slashed{D}) (\varphi^\dagger \cdot \ell) 
    \ +\ \frac{|y_E|^2}{M^4} (\bar{\ell}\cdot \varphi) (i \slashed{D})^3 (\varphi^\dagger \cdot \ell) 
    \ + \ \mathcal{O}(1/M^6)\;.
\end{align}
The higher order operators are self-Hermitian. Note also that $(\varphi^\dagger \cdot \ell_L)$ 
is a $SU(2)_L$-singlet with hypercharge, $Y=-1$. Hence, in the SM EFT, we have the appearance of a
massless composite particle with the same quantum number as $E_L$. It is tempting to think that this
particle potentially creates an extra chiral anomaly piece in SM EFT. This can be seen explicitly
by expanding the dimension-6 operator in the broken, $SU(2)_L\times U(1)_Y \to U(1)_{\rm em}$, phase of this model 
\begin{align}
    \frac{|y_E|^2}{M^2} (\bar{\ell}\cdot \varphi) (i \slashed{D}) (\varphi^\dagger \cdot \ell) =
    \frac{|y_E|^2}{M^2} \biggl [ 
     |\varphi_0|^2 \bar{e}_L i \slashed{\partial} e_L + g' |\varphi_0|^2 (\bar{e}_L \gamma^\mu e_L) B_\mu 
    + (\bar{e}_L \gamma^\mu e_L) (\varphi_0 i\partial_\mu \varphi_0^*) \biggr ] + \dots \;, \label{eq:smeftbroken}
\end{align}
where the dots are neutrino and charged Goldstone boson contributions.\footnote{We are mostly focusing in
neutral vector boson vertices in this work. Charged-current effects can be demonstrated 
similarly.}
The first term in the RHS of Eq.~\eqref{eq:smeftbroken} provides corrections to the 
kinetic term of the left-handed electron, $e_L$,  the 
second term couples the leptonic current of $e_L$ to the gauge-boson $B_\mu$ associated with the
hypercharge $U(1)_Y$ gauge-symmetry, and the last term couples the left-handed leptonic 
current to the Higgs current.\footnote{The neutral Higgs component $\varphi_0$ is defined above Eq.~\eqref{eq:Lm}.}
All these dimension-6 operators 
need to cooperate in the WIs of Eqs.~\eqref{eq:WI1}\eqref{eq:WI2} and \eqref{eq:WI3}.
As an aside, it should be noted that in our analysis, 
the equations of motion for the SM-fields are not utilized.

More specifically, in \eqref{eq:smeftbroken}, we have for the left-handed electron kinetic term,
\begin{equation}
    \mathcal{L}^{\rm Kin} = \bar{e}_L \left ( 1 + |\epsilon|^2 \right ) i \slashed{\partial} e_L \equiv 
    \bar{e}_L' i \slashed{\partial} e_L' \;,
\end{equation}
where $\epsilon$ is the same parameter as in the one 
defined in Eq.~\eqref{eq:De}. We therefore redefine the left-handed electrons as\footnote{An analogous 
field redefinition, but this time for the left-handed SM neutrinos, 
can be found in Ref.~\cite{Broncano:2002rw} for a heavy neutral fermion decoupling,
e.g., in the neutrino see-saw mechanism.}
\begin{equation}
    e_L' = \left ( 1 + |\epsilon|^2 \right )^{1/2} e_L \;.
\end{equation}
From the second term in \eqref{eq:smeftbroken} and in the notation of \eqref{eq:Lgauge}, the neutral currents read
\begin{align}
    J_Z^\mu &= \bar{e}_L' \gamma^\mu \biggl [ -\frac{1}{2} (1+|\epsilon|^2)^{-1} + s_w^2 \biggr ] e_L' + \bar{e}_R \gamma^\mu (s_w^2) e_R +\dots
   \;, \\[2mm]
    J_{\rm em} &= \bar{e}_L' \gamma^\mu (-1) e_L' + \bar{e}_R \gamma^\mu (-1) e_R + \dots \;,
\end{align}
where the dots refer to the rest of the SM-field unchanged currents w.r.t the SM.
In the more suggestive notation of Appendix~\ref{app:A}, we may
separate  SM-couplings, $\widehat{a}_V^{(f)}$ and $\widehat{b}_V^{(f)}$, with the $(d=6)$ SM EFT ones 
$\widetilde{a}_V^{(f)}$ and $\widetilde{b}_V^{(f)}$, as
\begin{align}
    a_Z^{(e')} = \frac{g_Z}{2} \biggl [ -\frac{1}{2} (1+|\epsilon|^2)^{-1} + 2 s_w^2 \biggr ] 
    \equiv \widehat{a}_Z^{(e')} + \widetilde{a}_Z^{(e')}\;, 
    \qquad b_Z^{(e')} = \frac{g_Z}{4}(1+|\epsilon|^2)^{-1}  \equiv  
    \widehat{b}_Z^{(e')} + \widetilde{b}_Z^{(e')}\;,
\end{align}
where 
\begin{eqnarray}
    \widehat{a}_Z^{(e')} = \frac{g_Z}{2}\biggr [-\frac{1}{2} + 2 s_w^2 \biggr ]\;, \quad
    \widehat{b}_Z^{(e')} = \frac{g_Z}{4}\;, \quad \widetilde{a}_Z^{(e')} = - \widetilde{b}_Z^{(e')} = 
    \frac{g_Z}{4} \biggl [ \frac{|\epsilon|^2}{1+|\epsilon|^2} \biggr ] \;.
    \label{eq:ab}
\end{eqnarray}
By comparing with the corresponding currents in Eqs.~\eqref{eq:Jz} and \eqref{eq:Jem} 
for $e_L'$ in full theory, we observe that they are identical 
[recall Eq.~\eqref{eq:UeWe}: $|(U_e)_{11}|^2 = (1+|\epsilon|^2)^{-1}$]. Similarly, the 
Wilson coefficients of Eq.~\eqref{eq:Cphil} are also confirmed by just comparing the
$Zee$ and $Z\nu\nu$ generic dimension-6 vertices calculated in Appendix~\ref{app:A}
with those in Eq.~\eqref{eq:ab} expanded at $\mathcal{O}(|\epsilon|^2$).
However, to cancel chiral anomalies, we need the effect from the heavy $E$-particles. How does the
SM EFT reproduce their effects? 

The Goldstone-boson $G^0$, couples to the divergent current of the symmetry that is being broken.
The last term of Eq.~\eqref{eq:smeftbroken} results in 
\begin{equation}
 \frac{|y_E|^2}{M^2} (\bar{e}_L \gamma^\mu e_L) (\varphi_0 i\partial_\mu \varphi_0^*) \supset -\frac{i}{v}\, \left [\frac{|\epsilon|^2}{1+|\epsilon|^2} \right ] \, \bar{e}_L' \gamma^\mu e_L' (i \partial_\mu G^0) \;.
\end{equation}
The term inside the square bracket is the $|(U_e)_{12}|^2$, the same that appears in the heavy electron $E_L'$ contribution to the $Z$-current in full theory [see Eq.~\eqref{eq:ZEb}].
The corresponding Feynman rule (in the limit of zero electron mass, i.e., $y_e\to 0$) is 
\vspace{0.1cm}
\begin{center}
\begin{fmffile}{G0}
  \parbox{20mm}{\begin{fmfgraph*}(40,15)
    \fmfstraight
    \fmfleft{i1}
    \fmfright{o1,o2}
    \fmf{dashes,label=$\rightarrow q$}{i1,v1}
    \fmf{fermion,tension=0.3}{o1,v1}
    \fmf{fermion,tension=0.3}{v1,o2}
        \fmflabel{$G^0$}{i1}
    \fmflabel{$e'$}{o1}
        \fmflabel{$e'$}{o2}
\end{fmfgraph*}}
{$ \ = \ \frac{1}{v} \, \left [\frac{|\epsilon|^2}{1+|\epsilon|^2} \right ] \, \slashed{q} \, P_L \ \stackrel{\eqref{eq:ab}}{=} \
\frac{1}{M_Z}\, \slashed{q}\, (\widetilde{a}_Z^{(e')} + \widetilde{b}_Z^{(e')} \gamma^5 )\;. $ }
\end{fmffile}
\end{center}
\vspace{0.3cm}
It is important to notice that, when there is a SM EFT insertion in the $Z$-boson vertex, that is 
$\widetilde{a}_V^{(f)}\ne 0$ and $\widetilde{b}_V^{(f)}\ne 0$, there is always 
a Goldstone-boson coupled to the divergence of the spontaneously broken current with 
identical SM EFT insertions. In other words,  $Z$-bosons are longitudinally polarized
when SM EFT insertions\footnote{That is, when one or more of the Wilson coefficients,
accompanying the operators in \eqref{eq:dim6ops},
are non-zero.} exist in a $Zff$-vertex.
If we call, $\Gamma^\mu_{\widetilde{Z}}$ ($\Gamma_{G^0}$) the dimension-6 SM EFT  
coupling of $Z$-boson (Goldstone-boson) to $e'$, then obviously the identity,
\begin{equation}
    q_\mu \Gamma^\mu_{\widetilde{Z}e'e'} + i M_Z \Gamma_{G^0e'e'} = 0 \;,
\end{equation}
holds. Hence, when  we take the divergence of the $Z$-boson current, dimension-6 SM EFT effects cancel out, leaving 
behind only the SM $Z$-boson coupling to $e'$, that is $\Gamma_{\widehat{Z}}$,
\begin{equation}
   q_\mu \Gamma^\mu_{{Z}e'e'} + i M_Z \Gamma_{G^0e'e'} = q_\mu \Gamma^\mu_{\widehat{Z}e'e'} \;. 
   \label{eq:GammaZG0}
\end{equation}
One may check this identity in general SM EFT in broken phase by applying the Feynman Rules of Appendix~\ref{app:A} (or those of Ref.~\cite{dedes:2017zog}). 
Eq.~\eqref{eq:GammaZG0} holds for a generic dimension-6 truncation in the SM EFT
and we have made a great use of it in section~\ref{sec:triangles}. 
The above results for the SM EFT in this particular model verifies all the ingredients 
used in the bottom-up approach. From now on, the cancellation
of chiral anomalies repeats itself exactly as presented in section~\ref{sec:AnomSMEFT}.

\section{A little phenomenology}
\label{sec:pheno}

In this section, we present numerical results for the form factor $h_3^\gamma(s)$ in the 
SM, in the UV-theory of section~\ref{sec:Zgg} and its SM EFT [sections~\ref{sec:ggZZgZ} and~\ref{sec:SMEFT}],
where a heavy vector-like electron $E$ of mass $M=1$ TeV  is decoupled. 
As we have seen,
corrections to $h_3^\gamma(s)$ relative to the SM are of the form $y_E^2 v^2/M^2$
and may be significant for large $y_E$ in the energy region between the $Z-$boson mass and the top mass. 
Furthermore, in the high-energy region,
these effects are governed by the dominance of dimension-6 terms, $(\frac{M_Z^2}{s})$,
with respect  to dimension-8 terms of the form $\frac{M_Z^4}{M^4}$. 

\begin{figure}[t]
    \centering
    \includegraphics[width=0.7\linewidth]{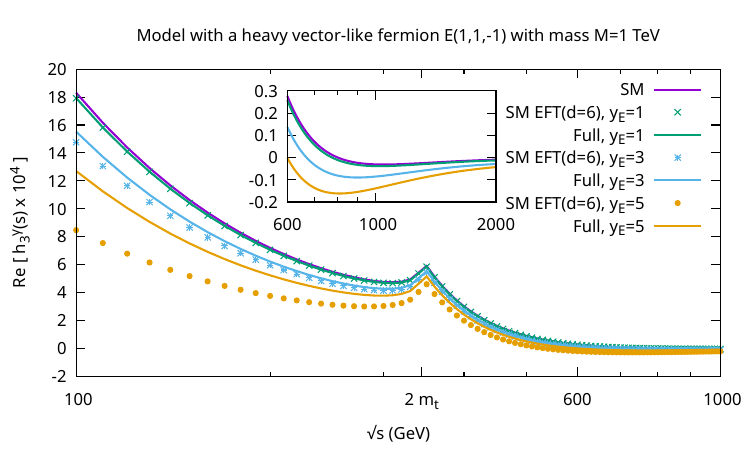}
    \caption{\sl The real part of the nTGV $\gamma^*\gamma Z$ form factor 
    ${\rm Re}\:[h_3^\gamma(s)\times 10^4]$ vs. the centre of mass energy $\sqrt{s}$. Solid (purple)
    lines refer to the SM   and, the full theory with a heavy vectorlike lepton $E(1,1,-1)$ of mass $M=1$ TeV,  with Yukawa coupling, $y_E=1$ (green), $y_E=3$ (cyan) and $y_E=5$ (orange). The respective dimension-6 SM EFT predictions are presented with  (green) crosses, (cyan) stars  and (orange) dots. 
    The inset graph focuses in the 
    high-energy region and displays only the full theory results (same axes and notation
    as in the outset graph but with $\sqrt{s} \in [600,2000]$ GeV). }
    \label{fig:h3g}
\end{figure}

These features are depicted in Fig.~\ref{fig:h3g} where the real part of $h_3^\gamma(s)$ 
versus the centre of mass energy $\sqrt{s}$, for various values of the Yukawa coupling, $y_E$,
is calculated numerically and plotted. 
%
The input value  $y_E=1$ is taken according to the maximally allowed by the  global SM EFT fit~\cite{Celada:2024mcf} and in particular from electroweak precision observables. 
The bounds on (assumed flavour universal) 
Wilson Coefficients $C^{\varphi\ell(1,3)}$,
and hence on the Yukawa coupling through Eq.~\eqref{eq:Cphil}, is 
$y_E \lesssim 1.3$. The input value $y_E=3$ is taken as almost the twice this bound
just as a theoretical uncertainty and $y_E=5$ for investigating the EFT validity 
versus the full theory results, although is experimentally excluded. 
The mass of the heavy fermion is $M=1$ TeV everywhere.

For two reference values of $\sqrt{s}$, 
the SM prediction is\footnote{We have set all light lepton and quark masses to zero. 
Their effects in $h_3^\gamma(s)$ for $\sqrt{s}\gtrsim M_Z$ are negligible .} 
\begin{equation}
\mathrm{SM:}\quad  
h_3^\gamma(\sqrt{s}=200~\mathrm{GeV}) = 7.2\times 10^{-4}\;, \qquad 
h_3^\gamma(\sqrt{s}=900~\mathrm{GeV}) = -2.4 \times 10^{-6} -3.0 \times 10^{-5} i \;, 
\label{eq:SMh3g}
\end{equation}
in agreement with the results of Ref.~\cite{Gounaris:2000tb}.
The heavy fermion-$E$ decoupling in full theory results in 
changing $h_3^\gamma(s)$ relative to the SM, as
%
\begin{align}
&(y_E=1):  
h_3^\gamma(\sqrt{s}=200~\mathrm{GeV}) = 7.1\times 10^{-4}\;, \quad 
h_3^\gamma(\sqrt{s}=900~\mathrm{GeV}) = -3.3 \times 10^{-6} -3.0 \times 10^{-5} i \;, \nonumber \\[2mm]
&(y_E=3):  
h_3^\gamma(\sqrt{s}=200~\mathrm{GeV}) = 6.2\times 10^{-4}\;, \quad 
h_3^\gamma(\sqrt{s}=900~\mathrm{GeV}) = -8.9 \times 10^{-6} -3.0 \times 10^{-5} i \;.
\label{eq:fullhg3}
\end{align}
%
There is a reduction of about 2\% for $y_E=1$ (15\% for $y_E=3$) for $\sqrt{s}=200$ GeV 
because of the negative Wilson coefficient [see \eqref{eq:Cphil}]
inserted in Eq.~\eqref{eq:h3l}. This is a particular effect of the fermion  
$E$ being decoupled  and 
is not generic to all gauge group representations of possible heavy fermions. 
Visible in Fig.~\ref{fig:h3g} is the top-mass threshold peak, 
where the imaginary part of $h_3^\gamma$ receives large values and tends to vanish asymptotically 
 $\sqrt{s} \gg 2 m_t$, as expressed by the second line of Eq.~\eqref{eq:h3glarges}.

As shown in the inset plot of Fig.~\ref{fig:h3g}, 
at energies near  the high  mass $M=1$ TeV,  the form factor $h_3^\gamma(s)$ 
takes on  smaller values, which peak at $\approx 10^{-6} (10^{-5})$ for $y_E=1$ ($y_E=3$) and are 
about a factor of 1.4 (4) bigger (in absolute value) than the SM expectation 
e.g., compare \eqref{eq:fullhg3} and~\eqref{eq:SMh3g} with $\sqrt{s}=900$ GeV.
As mentioned previously, in this region the SM drops like $1/s^2$ whereas the 
full theory (and also the SM EFT) like $1/s$.

The SM EFT of this model truncated at the dimension-6 level shows good agreement 
for Yukawa couplings with $y_E \lesssim 3$ compared to the full theory (cyan solid line in Fig.~\ref{fig:h3g}).
For larger values, i.e. $y_E =5$,  the dimension-8 and higher SM EFT contributions become important. However,
in the latter case, the perturbation parameter of the theory in \eqref{eq:Ueexpansion}, 
$\epsilon \simeq 0.9$, signals the EFT-expansion breakdown\footnote{In this case the $Z$-boson 
couples equally or even stronger to heavy  $(E)$ than to light  $(e)$ electrons,
invalidating the EFT concept in general.} if not the perturbative unitarity 
of the full model.


The first experimental search on neutral trilinear gauge boson vertices\footnote{Sometimes referred to as 
anomalous Trilinear Gauge boson Couplings (aTGCs).} was performed at LEP through the precesses 
$e^+e^- \to Z\gamma$ with the $Z$-boson decaying to either quarks, charged leptons or neutrinos~\cite{OPAL:2000ijr,L3:2004hlr,DELPHI:2007gzg}.
The bound set at $|h_3^\gamma(s)|<0.05$ in the region $\sqrt{s}\approx 200$ GeV was
about two orders of magnitude above the SM prediction in \eqref{eq:SMh3g}. 
Tevatron searches~\cite{D0:2009olv,CDF:2011rqn} for nTGVs bypassed by the LHC ones~\cite{CMS:2016cbq,ATLAS:2018nci,ATLAS:2019gey}. 
The process under experimental scrutiny is $pp\to Z(\nu\bar{\nu})\gamma$
with a high-energy photon, $E^\gamma > 600$ GeV, sets a bound around $|h_3^\gamma(s)|\lesssim 4 \times 10^{-4}$.
For reasons not clear to the authors, analyses at ATLAS and CMS treat $h_3^\gamma(s)$ as a static parameter.
Hence, with some degree of uncertainty, we can assert that the LHC bound is approximately two orders of magnitude away from the SM prediction~\eqref{eq:SMh3g} for $\sqrt{s}=900$ GeV. Nonetheless, for $y_E=3$, it narrows by an order of magnitude in the presence of a heavy vector-like electron in proximity, as illustrated in \eqref{eq:fullhg3}.

If the sensitivity of  future $e^+e^-$-colliders for $\Re e [h_3^\gamma(s)]$ and $\sqrt{s} = 200$ GeV is going to approximately be $\sim 10^{-3}$~\cite{Liu:2024tcz}, then the SM prediction will be at the threshold of being measured. Almost certainly, the
working model with a heavy electron is impossible to be searched for indirectly, since, 
as Fig.~\ref{fig:h3g} suggests, there is a reduction of about 15\% on $h_3^\gamma$ for $y_E=3$.
The question is if there are BSM effects from other than $E$ \textit{single} vector-like fermion models 
that may push this value above $10^{-3}$. The answer is given easily by the SM EFT prediction 
in Eq.~\eqref{eq:h3l} combined with the dictionary of Wilson coefficients at tree level of Ref.~\cite{deBlas:2017xtg}. We find that enhanced predictions for $h^\gamma_3$ relative to the SM
appear in  models with heavy vector-like 
leptons, $\Sigma (1,3)_0$, $\Delta_3 (1,2)_{-3/2}$
and heavy vector-like quarks, $T_2(3,3)_{2/3}, Q_5(3,2)_{-5/6}$. A similar analysis can be performed in 
the high-energy region based on our analytic formula \eqref{eq:h3glarges}.

It is clear that in order to have appreciable nTGVs the corresponding Yukawa coupling $y_E$
has to be large.  This raises questions, about  bounds from electroweak 
precision observables~\cite{Lavoura:1992np} since
the $Z$-boson coupling to light fermions must be substantially modified.
Needless to mention, of course, direct LHC searches for vector-like quarks and leptons~\cite{ATLAS:2023sbu,ATLAS:2024fdw}. 
A detailed phenomenological analysis, however, goes beyond the scope of our study. 

\subsection{Other UV-models}

The simple UV-theory with a single vector-like fermion we described, although complete, is not the end of 
the story. There may be mass admixtures of multiple heavy fermions with the SM Higgs~\cite{deBlas:2017xtg,Cepedello:2024ogz,Ellis:2024omd}. Such models avoid the heavy-light (SM) particle
mixing, and therefore constraints from electroweak observables are relaxed.
Related to the UV-model with the single heavy vector-like fermion $E(1,1,-1)$
of the previous section, one may envisage a model with an extra heavy vector-like doublet $L(1,2,-1/2)$ with
\begin{align}
    \mathcal{L}_{\mathrm{BSM}} = \mathcal{L}_\mathrm{SM} + \bar{E}i \slashed{D} E + \bar{L} i \slashed{D} L 
    -M_E \bar{E} E - M_L \bar{L} L 
    -y_R \bar{L}_L \cdot \varphi E_R - y_L \bar{L}_R \cdot \varphi E_L + \mathrm{h.c.} \;,
\end{align}
where Yukawa couplings involving heavy-light fields, 
like the $y_E \bar{\ell}_L\cdot \varphi E_R$ studied before, are
now set to zero. After electroweak symmetry breaking, the two heavy fields $L$ and $E$ mix each other
with squared mass difference proportional to $M v (y_R+y_L)$. In addition,  vector-axial couplings 
to the $Z$-boson, $b_Z^{(E)}$ and, $b_Z^{(L)} \sim (y_R-y_L)$, are produced. All chiral anomalies in this model 
cancel as expected, and a calculation of $h_3^\gamma(s)$ in the full theory at high-energies yields
\begin{equation}
    h_3^\gamma(s) = \frac{e}{\pi^2} \biggl [ -\frac{3}{4} \frac{M_Z^4 \, y_t^2}{s^2} \biggl (\ln^2\frac{s}{m_t^2}-\pi^2 \biggr ) \ + \ \frac{M_Z^4}{96 M^4} (y_R^2 - y_L^2) \ + \ \dots \biggr ]\;, \quad m_t^2 \ll s \ll M^2 \;,
\end{equation}
where we have taken $M_E=M_L\equiv M$ and $y_t$ is the top-Yukawa coupling. For the unnatural\footnote{Rather than the custodial symmetry supported blind spot, $y_L=y_R$.} 
choice $y_L\approx 0$ and $y_R \approx y_t$, 
the correction due to heavy leptons arises from dimension-8 operators in agreement with Refs.~\cite{Cepedello:2024ogz,Ellis:2024omd}, but its effect is, in the best case, 
less than one per mile  of the SM contribution for $M=1$ TeV.

Furthermore, there may be  additional
heavy Higgs fields~\cite{Adhikary:2024esf}, 
or even anomaly free sets of chiral fermions that receive masses from the SM Higgs boson,
although the latter scenarios are highly disfavoured by Higgs searches~\cite{Barducci:2023zml}.
Other UV-complete models, not discussed in this article, may contain additional heavy vector bosons 
coupled to the SM (or even to extra) fermions. Then phenomenologically interesting phenomena
can be handled with our analysis in section~\ref{sec:4fermionVs}, for instance Eq.~\eqref{eq:4fD}.
We leave quantitative results for all nTGV and related UV-models to a future work.

\section{Conclusions}
\label{sec:conclusions}

   We have demonstrated the chiral anomalies in 
   the Standard Model Effective Field Theory (SM EFT) expansion with multiple dimension-6 
   insertions of operators are cancelled, in the sense that the classical WIs [Eqs.~\eqref{eq:WI1}\eqref{eq:WI2} and
   \eqref{eq:WI3}] can always be restored in one-loop triangle diagrams. This cancellation is achieved in broken 
   electroweak phase through a careful choice of loop momentum routing in triangle diagrams, depicted in Fig.~\ref{fig:2} and Table~\ref{tab:1}, involving higher-dimensional operators. The key to this result lies in the fact that the Standard Model (SM) itself is anomaly-free [see Eq.~\eqref{eq:afree}], and the Goldstone boson contributions in the SM EFT ensure that gauge invariance and Bose symmetry are preserved. This cancellation holds for each SM fermion circulating in the one-loop triangle diagrams, regardless of the external gauge bosons involved.

   We have calculated the $C$-odd and $P$-odd form factors for neutral triple gauge boson vertices (nTGVs), such as $Z^{*}\gamma\gamma$, $V^{*}\gamma Z$, and $V^{*}ZZ$, in the presence of dimension-6 operators. 
   The master equation for those vertices is Eq.~\eqref{eq:mastervertex}. These vertices, which do not exist at tree level in the SM EFT, receive finite one-loop corrections that are sensitive to the effects of higher-dimensional operators. 
   
   The form factors for nTGVs, such as $h_{3}^{V}(s)$ and $f_{5}^{V}(s)$, exhibit distinct behaviours at different energy scales. At low energies (below the top-quark threshold), the SM contributions dominate, but the effects of dimension-6 operators, \eqref{eq:dim6ops}, modify the prediction, e.g., of $h_3^\gamma(s)$, by at most 
   an amount of approximately 2\%  
   for perturbative and experimentally allowed Yukawa coupling $y_E=1$ and $\sqrt{s}=200~\mathrm{GeV}$. Above the top-quark threshold, these form factors scale like $1/s$, rather than $1/s^2$ in the SM, reflecting the non-decoupling of heavy fermions from the UV-theory and may have a significant impact on $h_3^\gamma$ (up-to a factor of 1.4 w.r.t the SM). 
   This behaviour, is consistent with the expectations from effective field theory, as it is shown in Fig.~\ref{fig:h3g}, and provides a clear signature for new physics beyond the SM.

   We have examined a specific UV-complete model involving a heavy vector-like electron to validate our bottom-up approach. By integrating out the heavy fermion, we derived the corresponding SM EFT and confirmed that the chiral anomaly cancellation and the form factor $h_3^\gamma(s)$  are consistent with the full theory [see Eqs.~\eqref{eq:h3gbelowM} and \eqref{eq:h3glfull}]. This top-down approach highlights the importance of including heavy fermion effects in the SM EFT and provides a concrete example of how UV physics manifests in the low-energy effective theory. Our analysis strongly suggests that dimension-6 operators have a more significant impact than dimension-8 (or higher) operators both at low energies near and above the $Z$-boson mass,
   as well at high energies near the heavy mass. This makes them potentially observable in future collider experiments.

In summary, this work establishes the consistency of chiral anomaly cancellation in the SM EFT, provides a detailed analysis of nTGVs, and highlights the potential for observing new physics effects in future collider experiments. The interplay between theoretical consistency and phenomenological observability underscores the importance of effective field theory in exploring the frontiers of particle physics.

\section*{Acknowledgements}
We would like to thank Janusz Rosiek and Kristaq Suxho for enlightening blackboard discussions 
and critical comments on the manuscript. AD would 
like to thank Pascal Anastasopoulos and Elias Kiritsis  for various discussions.
AD acknowledges support from the COMETA COST Action CA22130.

\appendix
\section{Vector and Goldstone boson couplings to fermions in  SM EFT}
\label{app:A}

Explicit results for the chiral anomalies and triple-neutral gauge-boson-vertices appearing in sections~\ref{sec:AnomSMEFT} and~\ref{sec:nTGC} respectively,
need as inputs the vector- and Goldstone-boson coupling to fermions in the SM EFT with the $d=6$ operators 
arranged in Eq.~\eqref{eq:dim6ops}.
These have been found in Refs.~\cite{dedes:2017zog,Dedes:2023zws} and stated below for complementarity.
The  Feynman rules for the $V\bar{f}f$-vertex are $ - i \gamma^\mu (a_V^{(f)} + b_V^{(f)} \gamma^5)$
with ${a}_V^{(f)}\equiv  \widehat{a}_V^{(f)}+\widetilde{a}_V^{(f)}$ and ${b}_V^{(f)}\equiv  \widehat{b}_V^{(f)}+\widetilde{b}_V^{(f)}$ being the “SM-like” plus the “EFT-like” parts. Then
for $V=Z$ we obtain\footnote{We do not include the dipole ($\sigma^{\mu\nu}$) parts in the following expressions.}
\begin{align}
\widehat{a}_Z^{(\nu)} &= \frac{1}{4} g_Z\;,\quad  
\widetilde{a}_Z^{(\nu)}=- \frac{1}{4} g_Z \frac{v^2}{\Lambda^2}\,  ( C^{\varphi \ell (1)}-C^{\varphi \ell (3)})\;, \label{eq:nunuZ}\\
\widehat{b}_Z^{(\nu)} &=-\frac{1}{4} g_Z\;, \quad \widetilde{b}_Z^{(\nu)}= \frac{1}{4} g_Z \frac{v^2}{\Lambda^2}\,  ( C^{\varphi \ell (1)}-C^{\varphi \ell (3)})\;,  
\\[4mm]
\widehat{a}_Z^{(e)} &= -\frac{1}{4} (3 g' Z_{g'}Z_{\gamma Z}^{21}  + g Z_g Z_{\gamma Z}^{11})\;,\quad  
\widetilde{a}_Z^{(e)}=- \frac{1}{4} g_Z \frac{v^2}{\Lambda^2}\,  ( C^{\varphi e} + C^{\varphi \ell (1)}+C^{\varphi \ell (3)})\;, \\
\widehat{b}_Z^{(e)} &=\frac{1}{4} g_Z\;, \quad \widetilde{b}_Z^{(e)}= - \frac{1}{4} g_Z \frac{v^2}{\Lambda^2}\,  ( C^{\varphi e} - C^{\varphi \ell (1)}-C^{\varphi \ell (3)})\;, 
\\[4mm]
\widehat{a}_Z^{(u)} &= \frac{1}{12} (5 g' Z_{g'}Z_{\gamma Z}^{21}  + 3 g Z_g Z_{\gamma Z}^{11})\;,\quad  
\widetilde{a}_Z^{(u)}=- \frac{1}{4} g_Z \frac{v^2}{\Lambda^2}\,  ( C^{\varphi u} + C^{\varphi q (1)}-C^{\varphi q (3)})\;, \\
\widehat{b}_Z^{(u)} &= - \frac{1}{4} g_Z\;, \quad \widetilde{b}_Z^{(u)}= - \frac{1}{4} g_Z \frac{v^2}{\Lambda^2}\,  ( C^{\varphi u} - C^{\varphi q (1)}+C^{\varphi q (3)})\;, 
\\[4mm]
\widehat{a}_Z^{(d)} &= -\frac{1}{12} ( g' Z_{g'}Z_{\gamma Z}^{21}  + 3 g Z_g Z_{\gamma Z}^{11})\;,\quad  
\widetilde{a}_Z^{(d)}=- \frac{1}{4} g_Z \frac{v^2}{\Lambda^2}\,  ( C^{\varphi d} + C^{\varphi q (1)}+C^{\varphi q (3)})\;, \\
\widehat{b}_Z^{(d)} &=  \frac{1}{4} g_Z\;, \quad \widetilde{b}_Z^{(d)}= - \frac{1}{4} g_Z \frac{v^2}{\Lambda^2}\,  ( C^{\varphi d} - C^{\varphi q (1)}-C^{\varphi q (3)})\;, 
\end{align}
and for $V=\gamma$,
\begin{align}
    \widehat{a}_\gamma^{(\nu)} = 0\;, \quad \widehat{a}_\gamma^{(e)} = - g' Z_{g'} Z_{\gamma Z}^{22}\;, \quad
    \widehat{a}_\gamma^{(u)} = \frac{2}{3} g' Z_{g'} Z_{\gamma Z}^{22}\;, \quad
    \widehat{a}_\gamma^{(d)} = -\frac{1}{3} g' Z_{g'} Z_{\gamma Z}^{22}\;.\label{eq:gammaff}
\end{align}
The coupling $g_Z=-g' Z_{g'} Z_{\gamma Z}^{21} + g Z_g Z_{\gamma Z}^{11}$, 
the normalization factors $Z_{g'},Z_g$ and the matrix $Z_{\gamma Z}$
can be stated as such to all orders in the EFT expansion. Their explicit expressions are given
in Ref.~\cite{Dedes:2023zws} up to dimension-8 operators in the SM EFT. However, chiral-anomalies with 
“SM-like” couplings $\widehat{a}_V^{(f)}$ and $\widehat{b}_V^{(f)}$ in Eq.~\eqref{eq:afree} cancel  without any
reference to their specific mathematical expressions. 

In this notation, the Feynman rules for the Goldstone-boson couplings to fermions, with incoming momentum $q$ in the Goldstone-boson line, obtain a particularly simple form:
\begin{align}
G^0(q)\bar{\nu}\nu : &\qquad  -\frac{m_e}{v Z_{G^0}} \gamma^5+  \frac{\slashed{q}}{M_Z}\, (\widetilde{a}_Z^{(\nu)} + \widetilde{b}_Z^{(\nu)}\gamma^5)\;, \label{eq:G0nunu}\\
G^0(q)\bar{e}e : &\qquad  \frac{m_e}{v Z_{G^0}} \gamma^5 + \frac{\slashed{q}}{M_Z}\, (\widetilde{a}_Z^{(e)} + \widetilde{b}_Z^{(e)}\gamma^5)\;, \\
G^0(q)\bar{u}u : &\qquad  -\frac{m_u}{v Z_{G^0}} \gamma^5 + \frac{\slashed{q}}{M_Z}\, (\widetilde{a}_Z^{(u)} + \widetilde{b}_Z^{(u)}\gamma^5)\;, \\
G^0(q)\bar{d}d : &\qquad  \frac{m_d}{v Z_{G^0}} \gamma^5 + \frac{\slashed{q}}{M_Z}\, (\widetilde{a}_Z^{(d)} + \widetilde{b}_Z^{(d)}\gamma^5)\;.\label{eq:G0dd}
\end{align}
The normalization factor $Z_{G^0}$ cancels in the calculation of anomalies, as noted
in Eq.~\eqref{eq:trigonoG}, and $M_Z=\frac{1}{2} g_Z v Z_{G^0}$.
It is crucial for the reader to notice the similarity between the SM EFT part of  
$Z\bar{f}f$ and $G^0\bar{f}f$ vertices; exactly this feature has been specifically identified  in section~\ref{sec:SMEFT} for the decoupling of a heavy electron. 

Moreover, Eqs.~\eqref{eq:G0nunu}-\eqref{eq:G0dd} 
have been used in calculating the Goldstone-boson triangle diagram 
in Eq.~\eqref{eq:trigonoG}
where $\slashed{q}$ is factored out from 
the anomaly terms $\Delta^{\mu\nu\rho\, (f)}_{\widetilde{V}_iV_jV_k}$ with 
$\widetilde{V}_i$ containing either $\widetilde{a}_V^{(f)}$ or $\widetilde{b}_V^{(f)}$
of Eqs.~\eqref{eq:nunuZ}-\eqref{eq:gammaff}.

\bibliography{A-SMEFT}{}
\bibliographystyle{JHEP}

\end{document}